\documentclass[11pt]{article}

\usepackage{kotex}
\usepackage{graphicx}
\usepackage{array}
\usepackage{tabularx}
\usepackage{amsmath,amssymb,bbold}
\usepackage{url}
\usepackage{rotating}
\usepackage{color}
\usepackage{hhline}
\usepackage{hyperref}
\usepackage{slashed}
\usepackage{makecell}
\usepackage{empheq}
\usepackage{physics}
\usepackage{comment}
\usepackage{authblk}
\usepackage{xcolor}
\usepackage[dvips=false,pdftex=false,vtex=false]{geometry}
\usepackage{changepage}
\usepackage{simpler-wick}
\usepackage{makecell}
\usepackage{multirow}
\usepackage{cleveref}
\usepackage{cite}

\begin{document}

\begin{flushright}
KIAS-Q23006
\end{flushright}

\vskip 1.cm

\begin{center}
{\Large \bf All the matrix elements of
covariant tensor currents \\[5pt] 
of massless particles in the covariant formulation}\\[20pt]
{Jaehoon Jeong\footnote{jeong229@kias.re.kr}} \\[15pt]
{\it  Quantum Universe Center, KIAS, Seoul 02455, Korea}
\end{center}

\vskip 0.5cm

\begin{abstract}
We present an efficient algorithm for constructing all the matrix elements of covariant 
tensor currents of massless particles of arbitrary spins in the covariant formulation.
The construction of matrix elements can be taken simply by assembling 
the basic matrix elements which are derived from the basic three-point vertices.
We obtain the selection rules for the decay of an off-shell massive particle 
into two identical massless particles which are the generalization 
of the Landau-Yang (LY) theorem. After showing how to identify fully the discontinuity 
between the matrix elements for the spacelike and lightlike momentum transfers, 
we derive all the matrix elements of conserved 
tensor currents of massless particles including high spins, from which the Weinberg and Witten (WW) theorem is automatically extracted with additional limits on the particles.
\end{abstract}

\section{Introduction}

The Standard Model (SM) with the symmetry-broken electroweak 
sector~\cite{Glashow:1961tr} has been completed after the discovery 
of the Higgs boson at the Large Hadron Collider (LHC)~\cite{ATLAS:2012yve}. 
Despite the great success of the SM, we are now encountering various 
unsolved problems in the SM such as dark matter~\cite{Bertone:2004pz}, 
neutrino oscillation~\cite{Super-Kamiokande:1998kpq}, and matter-antimatter 
asymmetry~\cite{Dine:2003ax}. However, no new particles and phenomena 
have been observed so far at the LHC~\cite{ParticleDataGroup:2020ssz}. 
In such a situation with no decisive evidence for new physics beyond the SM, 
studying all the allowed effective interactions of any particles including 
high spins in a model-independent way can be one of the
powerful strategies for new physics searches.
\\

The high-spin massless particles have been hotly 
studied in terms of their theoretical structures described in various 
representations~\cite{Fierz:1939ix,Weinberg:1964ev}. In addition, 
the construction of the field theories allowing their appearance was taken 
with light-cone coordinates~\cite{deWit:1979sib} and also 
in the supersymmetry~\cite{Curtright:1979uz}. 
In contrast to the remarkable developments, their existence was 
threatened by several restrictions so-called ``no-go'' 
theorems~\cite{Landau:1948kw,Weinberg:1980kq,Porrati:2008rm,Weinberg:1964ew}. 
However, {\it it does not mean their complete absence in nature}: 
Landau and Yang have shown one of the no-go theorems which are the selection rules
on the decay of a particle into two spin-1 massless photons~\cite{Landau:1948kw} 
and it has been extended recently in terms of the decay 
for two identical massless particles of any spin~\cite{Choi:2021ewa},\footnote{
We notice that the partial extension of the LY theorem has been already taken in 
terms of spin-0 and spin-1/2 massless particles in Ref~\cite{Pleitez:2015cpa}.}  but
all the decay modes are not ruled out even by the generalization. 
Weinberg and Witten have presented another one of 
the no-go theorems~\cite{Weinberg:1980kq}\footnote{For 
review, see~\cite{Loebbert:2008zz}} 
that all theories allowing the construction of a covariant conserved current $J_{\mu}$ cannot 
carry massless particles of spin-$s>1/2$ with 
a non-vanishing expectation value of the conserved 
charge $\int d^3 \vec x \;J_{0}$ and all theories allowing 
the construction of an energy-momentum tensor $T_{\mu\nu}$ cannot 
carry massless particles of spin-$s>1$ with a non-vanishing expectation 
value of the four-momentum $\int d^3 \vec x \;T_{0\mu}$. However, in the SM~\cite{ParticleDataGroup:2020ssz}, we have already 
two spin-1 massless particles, photon and gluon, which do not 
carry a conserved charge and a covariant conserved current respectively. In addition, 
the discovery of gravitational waves~\cite{LIGOScientific:2016aoc} implies strongly 
the presence of their quanta, a spin-2 massless particle called the graviton.    
\\

In Ref.~\cite{Choi:2021qsb}, 
an efficient algorithm for constructing the general covariant three-point 
vertices including on-shell particles 
has been developed in the two-body decay by utilizing 
the closely-related two mathematical frameworks, 
the helicity formalism~\cite{Jacob:1959at,Wick:1962zz,Chung:1971ri} and 
covariant formulation, in the Jacob-Wick convention~\cite{Jacob:1959at}. 
However, its application is significantly limited because all the external particles are 
taken to be on-shell. To cover the three-point interactions even 
including {\it off-shell particles}, we extend and complement their algorithm to construct 
all the matrix elements of covariant tensor currents of 
massless particles of arbitrary spins in the covariant formulation. We 
in passing note that there are other powerful approaches in obtaining the covariant 
tensor currents~\cite{Chung:1997jn}.
\\

This algorithm enables us to achieve the followings:
We can obtain the selection rules for the decay of an off-shell massive particle 
into two identical massless particles. The off-shell particle includes the lower 
spin degrees of freedom as well as its intrinsic ones, thus we can obtain the 
more general selection rules including those for the on-shell particle 
cases~\cite{Landau:1948kw,Choi:2021ewa} automatically. The discontinuity 
of all the (1-to-1) matrix elements between one-particle states can be identified fully in 
this algorithm. After constructing all the covariant three-point vertices 
yielding the matrix elements of conserved tensor currents (including noncovariant ones)
\footnote{The noncovariant conserved current and energy-momentum tensor of a spin-3/2 massless particle was derived in Ref.~\cite{Sudarshan:1981cj}.},
we show that the WW theorem is extracted automatically with additional limits 
on massless particles.
\\

This work is organized as follows: In Sec.~\ref{sec:symmetric_currents}, 
we first discuss the validity 
of dealing with symmetric covariant tensor currents when they are coupled to 
the Feynman propagator~\cite{Huang:2005js,Hayashi:1969qb} of an integer-spin 
particle. Thereafter, in the decay of an integer-spin particle into two massless 
particles, we show how the corresponding 0-to-2 matrix elements are represented 
in the covariant formulation. In Sec.~\ref{sec:wave_tensors}, the conventional 
wave tensors of massless particles included in the 0-to-2 matrix elements are 
given on a helicity basis in the Wick convention~\cite{Wick:1962zz}. 
In Sec.~\ref{sec:basic_and_form-factor_operators}, we focus on 
obtaining the basic and form-factor operators which are the basic building blocks 
of the covariant three-point vertices. In this procedure, the basic matrix elements 
are derived by the product of the basic operators and wave tensors, of which an 
appropriate combination leads to the construction of the general matrix elements 
directly. Employing the one-to-one correspondence between the covariant three-point 
vertices and matrix elements, in Sec.~\ref{sec:covariant_three-point_vertices}, 
we concentrate on identifying only the 
covariant three-point vertices. After the identification, we obtain 
the selection rules on the decay of an off-shell particle 
into two identical massless particles. In Sec.~\ref{sec:conserved_tensor_currents}, 
After deriving all the 1-to-1 matrix elements from the 0-to-2 matrix elements by means of
the crossing symmetry, we investigate the discontinuity of all the basic matrix elements, 
enabling us to identify the discontinuity of all the 1-to-1 matrix elements fully. 
The introduction of alternative basic operators leads us to the construction of 
the covariant three-point vertices for all the conserved tensor currents. We show that 
the WW theorem can be extracted from the conserved vertices with additional limits 
on massless particles including high spins. In Appendix.~\ref{sec:explicit_expressions}, 
the explicit expressions of all the bosonic basic operators are given 
for ease of reference in terms of the spacelike and lightlike momentum transfers.
\\

\section{Symmetric covariant tensor currents of massless particles}
\label{sec:symmetric_currents}

In the present work, the matrix elements of covariant tensor currents of massless particles
are assumed to be coupled to the Feynman propagator defined 
in terms of a conventional free field $\Psi_{\mu_1\cdots \mu_J}$~\cite{Behrends:1957rup} 
of the particle $\Psi$. To identify the constraints on the matrix elements, 
it is necessary to investigate the properties of integer-spin propagators. 
In momentum space, the Feynman propagator of $\Psi$ is given by
\begin{align}
\Pi_{F}^{\mu_1\cdots \mu_J \nu_1 \cdots \nu_J}(p)=\int d^4 x\; e^{ip\cdot x} \;\langle 0| 
T\{ \Psi^{\mu_1\cdots\mu_J}(x)\Psi^{\dagger \nu_1\cdots\nu_J}(0)\}|0\rangle,
\label{eq:propagator}
\end{align}
in terms of the 2-point correlation function 
$\langle 0| T\{ \Psi^{\mu_1 \cdots \mu_J} \Psi^{\dagger \nu_1\cdots \nu_J}\} |0\rangle$ 
with the momentum $p$ where the free field 
$\Psi_{\mu_1\cdots \mu_J}$ is expressed as
\begin{align}
\Psi^{\mu_1\cdots \mu_J}(x)=\int \frac{d^3 \vec k}{(2\pi)^3\sqrt{2E}}\sum_{\lambda=-J}^{J}
\Big[\varepsilon^{\mu_1\cdots \mu_J}( k,\sigma)a( k,\sigma)e^{-ik\cdot x}
+\varepsilon^{*\mu_1\cdots \mu_J}(  k,\sigma)b^\dagger( k,\sigma)e^{ik\cdot x}\Big],
\end{align}
with the momentum 
$k=(E,\vec k\,)$ satisfying $E=\sqrt{\vec k^{\,2}+m^2}$ and helicity $\sigma$
in terms of  the annihilation and creation operators, $a$ and $b^\dagger$, of 
the on-shell bosonic particle $\Psi$ and its anti-particle $\bar{\Psi}$ 
where the bosonic wave tensor $\varepsilon^{\mu_1\cdots \mu_J}$ with 
its complex conjugation $\varepsilon^{*\mu_1\cdots \mu_J}$ is 
symmetric, traceless, and divergence-free for any helicity value $\sigma=-J,\cdots ,J$
\cite{Choi:2021qsb,Behrends:1957rup},
\begin{align}
\varepsilon_{\alpha\beta\mu_i\mu_j}\varepsilon^{\mu_1\cdots \mu_i\cdots \mu_j\cdots \mu_J}
(k,\sigma)&=0,
\label{eq:symmetric}
\\[3pt]
g_{\mu_i\mu_j}\varepsilon^{\mu_1\cdots \mu_i\cdots \mu_j\cdots \mu_J}(k,\sigma)&=0,
\label{eq:traceless}
\\[3pt]
k_{\mu_i}\varepsilon^{\mu_1\cdots \mu_i\cdots \mu_J}(k,\sigma)&=0,
\label{eq:divergence-free}
\end{align}
satisfying the on-shell condition $(k^2-m^2)\varepsilon^{\mu_1\cdots \mu_J}=0$. 
\\

Note that the presence of the step functions involved by 
the time-ordered operation in the correlation function in 
Eq.~\eqref{eq:propagator} interrupts the propagator to be 
covariant.\footnote{The non-covariance of the propagator can be checked 
by taking the Lorentz transformation to the propagator. For a spin-1 particle, 
one can get $\Lambda^{\rho}_{\;\;\,\mu} \Lambda^{\sigma}_{\;\;\,\nu} \Pi^{\mu\nu}(p') 
\neq \Pi^{\rho\sigma}(\Lambda p')$.}. 
The noncovariant terms appearing in the propagator can 
be canceled out by adding appropriate noncovariant contact 
terms in the interaction Hamiltonian~\cite{Weinberg:1964cn}. 
This cure has to be carried out if one requires the $S$ matrix to be 
Lorentz-invariant. The covariant 
propagator of $\Psi$ in momentum space 
can be written without any 
difficulty in eliminating nonphysical degrees of 
freedom\footnote{For a massless particle, the structure 
of the covariant propagator depends on the choice of gauge 
(see the issues in Ref.~\cite{Hayashi:1969qb}).} as
\begin{align}
\Pi_{C}^{\mu_1\cdots \mu_J \nu_1 \cdots \nu_J}(p)
=\sum^{J}_{\sigma=-J} \big[\varepsilon^{\mu_1\cdots \mu_J}
(p,\sigma)\varepsilon^{*\nu_1\cdots \nu_J}( p,\sigma)\big] 
\frac{i}{p^{2}-m^2+i\epsilon},
\label{eq:covariant_massive_propagator}
\end{align}
with the projection operator $\sum \varepsilon^{\mu_1\cdots \mu_J}
\varepsilon^{*\nu_1\cdots \nu_J}$. Here, The projection operator is given 
by the proper combination of the spin-1 projection operators $(g_{ab}-p_{a}p_{b}/m^2)$ 
which is divergence-free only for an on-shell $\Psi$ (see the explicit expressions of 
the projection operators of massive 
particles of arbitrary spins in Ref.~\cite{Huang:2005js}). For an off-shell massive $\Psi$, 
the projection operator follows only the symmetric property of a wave tensor 
in Eq.~\eqref{eq:symmetric}. In the present work, thus, we assume the covariant tensor 
current to be at least symmetric with no loss of generality. 
{\it Note that the symmetric property of a bosonic propagator 
is guaranteed automatically regardless of the mass due to its definition 
including the symmetric correlation functions in Eq.~\eqref{eq:propagator}
even though the invariance of the $S$ matrix is not required, 
i.e. without canceling out the non-covariant contact terms.}     

\section{Wave tensors}
\label{sec:wave_tensors}

\begin{figure}[ht!]
\centering
\includegraphics[scale=1.4]{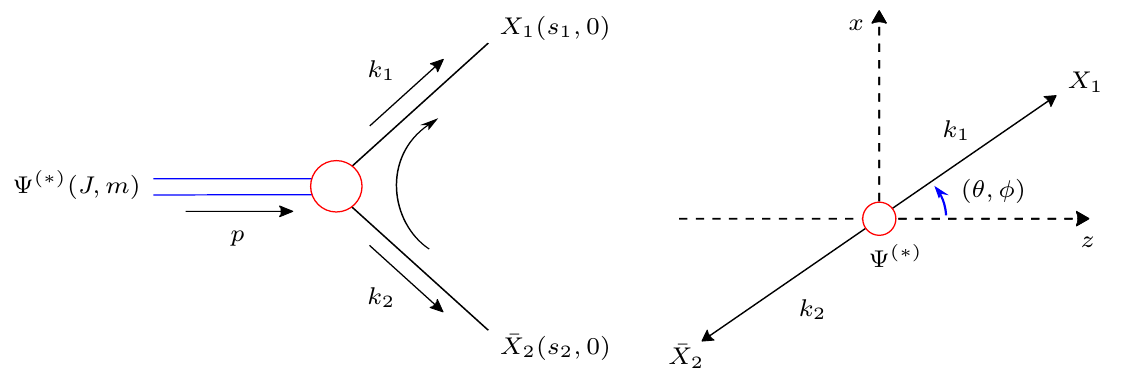}
\caption{A diagram and a kinematics description for the two-body decay 
$\Psi^{(*)} \rightarrow X_1 X_2$. (Left) A particle $\Psi^{(*)}$ of integer spin $J$ and mass $m$
decays into two massless particles, $X_1$ and $\bar{X}_2$ of spins, $s_1$ and $s_2$. 
The energy-momentum conservation determines the $\Psi$ momentum to be 
the summation $p=k_1+k_2$ of two momenta $k_{1,2}$ of $X_1$ and $\bar{X}_2$. 
The fermionic line of two massless particles is defined to flow from bottom to top. 
(Right) In the $\Psi^{(*)}$ rest frame ($\Psi$RF) with a fixed coordinate system, 
the two massless particles, $X_1$ and $\bar{X}_2$, 
with the momenta $k_{1,2}$ are moving in opposite directions to each other. 
$\theta$ and $\phi$ are the polar and azimuthal angles of the particle $X_1$.}
\label{fig:diagram_and_kinematic_configuration_decay}
\end{figure}

The algorithm for obtaining all the matrix elements of covariant tensor currents 
is developed in the decay of a particle $\Psi^{(*)}$ 
of integer spin $J$ and mass $m$ into two on-shell massless particles, 
$X_{1}$ of spin $s_1$ and $\bar{X}_2$ of spin $s_2$,
\begin{align}
\Psi^{(*)}(J,m) \rightarrow X_1(s_1,0) + \bar{X}_{2}(s_2,0),
\label{eq:two_body_decay}
\end{align}
where the symbol $*$ stands for the off-shell $\Psi^*$ and 
$\bar{X}_2$ is the anti-particle of $X_2$. The 0-to-2 matrix elements 
of the covariant tensor currents $\Theta^{[J;s_1,s_2]}_{\mu_1\cdots \mu_J}$, 
which are employed in obtaining the decay amplitudes, can be 
expressed on a helicity basis:
\begin{align}
\big({}_{1} \langle  k_1,\lambda_1| \otimes
{}_{\bar{2}} \langle  k_2,\lambda_2 |\big)\,\Theta^{[J;s_1,s_2]}_{\mu_1\cdots \mu_J}| \,0\, \rangle
&={}_{1,\bar{2}} \langle  k_1,\lambda_1; k_2,\lambda_2 |
\,\Theta^{[J;s_1,s_2]}_{\mu_1\cdots \mu_J}| \,0\, \rangle
\nonumber
\\
&\equiv{}_{1,\bar{2}}^{} \Theta^{[J;s_1,s_2]}_{(\lambda_1,\lambda_2)\mu_1\cdots \mu_J}(k_1,k_2),
\label{eq:definition_of_matrix_elements}
\end{align}
in terms of the momenta $k_{1,2}$ and helicities $\lambda_{1,2}$ of 
$X_1$ and $\bar{X}_2$ (see the diagram of the two-body decay on the 
left panel in Fig.~\ref{fig:diagram_and_kinematic_configuration_decay}). 
In helicity formalism~\cite{Jacob:1959at,Chung:1971ri,Wick:1962zz}, the helicity state of a particle is expressed conventionally with the labels of its momentum and helicity, e.g. $|k_{1,2},\lambda_{1,2} \rangle$ for $X_{1}$ and $\bar{X}_2$.  
In Sec.~\ref{sec:covariant_three-point_vertices}, however, they will be written more specifically 
with the subscripts which stand for the species of the particles, 
$X_{1}$ and $\bar{X}_2$, such that the 
conditions on the covariant matrix elements 
for two identical massless particles satisfying $X_{1}=\bar{X}_2$ can be 
derived without any confusion. The covariance of matrix elements enables us to find 
their explicit structures in a reference frame with simple calculations. 
\\

For the explicit construction, it is necessary to investigate 
the configurations of creation and 
annihilation operators of two massless particles in the covariant 
tensor currents because the anticommutativity between two fermionic state operators 
often involves a minus sign when coupled to the massless states. Thus, 
we assume that the covariant tensor currents include the following configurations,
\begin{align} 
\{b_1\, a_2,\; b_1\, b_2^\dagger,\; a_1^\dagger \,a_2,\; a_1^\dagger\, b_2^\dagger\} 
\in\Theta^{[J;s,s]}_{\mu_1\cdots \mu_J}, 
\label{eq:configuration_of_state_operators}
\end{align}
in terms of the state operators, $a_{1,2}$ and $b_{1,2}$, of the massless particles, 
$X_{1,2}$ and $\bar{X}_{1,2}$. Adopting the covariant formulation, 
we express the 0-to-2 matrix elements 
in Eq. \eqref{eq:definition_of_matrix_elements}
as a product of the wave tensors, $\bar{\phi}$ and $\phi$, of $X_{1}$ 
and $\bar{X}_{2}$ and a covariant three-point vertex $\Gamma$:
\begin{align}
\Theta^{[J;s_1,s_2]}_{(\lambda_1,\lambda_2)\mu_1\cdots \mu_J}(k_1,k_2)
&=\bar{\phi}^{\alpha_1\cdots\alpha_{n_1}}(k_1,\lambda_1)\,(-1)^{4s_1s_2}
\,\Gamma_{\alpha_1\cdots\alpha_{n_1},\beta_1\cdots\beta_{n_2};\mu_1\cdots \mu_J}
^{[J;s_1,s_2]}(k_1,k_2)\, 
\phi^{\beta_1\cdots\beta_{n_2}}(k_2,\lambda_2),
\label{eq:decay_matrix_elements}
\end{align}
in terms of the nonnegative integers $n_{1,2}=s_{1,2}$ or 
$n_{1,2}=s_{1,2}-1/2$ for the integer or half-integer spins $s_{1,2}$
with the phase $(-1)^{4s_1s_2}$ appearing due to the configurations 
in Eq.~\eqref{eq:configuration_of_state_operators}. 
As a massless particle can take only maximal-magnitude 
helicity values, the wave tensors $\phi_{1,2}$ can be written as the products of $s_{1,2}$ 
massless spin-1 polarization vectors respectively:
\begin{align}
\phi_1^{\alpha_1\cdots\alpha_{s_1}}(k_1,\pm s_1)
&=\varepsilon^{\alpha_1\cdots\alpha_{s_1}}(k_1,\pm s_1)
=\varepsilon^{\alpha_1}(k_1,\pm 1)\cdots
\varepsilon^{\alpha_{s_1}}(k_1,\pm 1)\,,
\label{eq:bosonic_wave_tensor1}
\\
\phi_2^{\beta_1\cdots\beta_{s_2}}(k_2,\pm s_2)
&=\varepsilon^{*\beta_1\cdots\beta_{s_2}}(k_2,\pm s_2)
=\varepsilon^{*\beta_1}(k_2,\pm 1)\cdots\varepsilon^{*\beta_{s_2}}(k_2,\pm 1)\,,
\label{eq:bosonic_wave_tensor2}
\end{align}
satisfying their own on-shell conditions, $k_1^2\, \varepsilon^{\alpha_1\cdots \alpha_{s_1}}
(k_1,\pm s_1)=0$ and $k^{2}_2\, \varepsilon^{\beta_1\cdots \beta_{s_2}}(k_2,\pm s_2)=0$. 
In the matrix elements in Eq. \eqref{eq:decay_matrix_elements}, the wave tensor 
$\bar{\phi}=\varepsilon^*$ of $X_1$ is the complex conjugation of 
$\varepsilon^{\alpha_1\cdots \alpha_{s_1}}(k_1,\lambda_1)$. 
Note that the bosonic wave tensors satisfy the three properties
in Eqs.~\eqref{eq:symmetric}, \eqref{eq:traceless}, and 
\eqref{eq:divergence-free},
regardless of the helicity values, $\lambda_1=\pm s_1$ and $\lambda_2=\pm s_2$, respectively. 
On the other hand, the fermionic wave spinors of the on-shell massless 
fermions, $X_{1}$ and $\bar{X}_2$, of half-integer spins $s_{1,2}$, are given by the products of 
massless spin-$n_{1,2}$ bosonic wave tensors with $n_{1,2}=s_{1,2}-1/2$ and spin-1/2 massless 
Dirac spinors, respectively:
\begin{align}
\phi^{\alpha_1\cdots \alpha_{n_1}}(k_1,\pm s_1)
&=u^{\alpha_1\cdots \alpha_{s_1}}(k_1,\pm s_1)
=\varepsilon^{\alpha_1}(k_1,\pm 1)\cdots\varepsilon^{\alpha_{n_1}}(k_1,\pm 1)
u(k_1, \pm\mbox{\small $\frac{1}{2}$}), 
\label{eq:fermionic_wave_tensor1}
\\
\phi^{\beta_{1}\cdots\beta_{n_2}}(k_2,\pm s_2)
&=v^{\beta_2\cdots \beta_{n_2}}(k_2,\pm s_2)
=\varepsilon^{*\beta_1}(k_2,\pm 1)\cdots\varepsilon^{*\beta_{n_2}}(k_2,\pm 1) 
v(k_2, \pm\mbox{\small $\frac{1}{2}$})\,,
\label{eq:fermionic_wave_tensor2}
\end{align}
where the Dirac spinors satisfy their own on-shell 
Dirac equations $\slash\!\!\! k_1\, u_1(k_1, \pm\frac12)=0$ and 
$\slash\!\!\! k_2\, v_2(k_2, \pm\frac12)=0$. In contrast to the bosonic case, 
the adjoint expression $\bar{\phi}$ of the $X_1$ wave tensor is given by 
$\bar{u}^{\alpha_1\cdots \alpha_{n_1}}=\gamma^0 (u^{\alpha_1\cdots \alpha_{n_1}})^\dagger$ with a gamma matrix $\gamma^0$. 
These fermionic wave spinors are not only symmetric, traceless, and 
divergence-free, but also $\gamma$ contraction-free relations,
\begin{align}
\gamma_{\alpha_i} u^{\alpha_1\cdots\alpha_i\cdots\alpha_{n_1}} (k_1,\lambda_1)
=\gamma_{\beta_j} v^{\beta_1\cdots\beta_j\cdots\beta_{n_2}} (k_2,\lambda_2)
=0,
\end{align}
in terms of arbitrary two nonnegative integers $i,j=1,\cdots, n_{1,2}$ 
regardless of the helicity values, $\lambda_1=\pm s_{1}$ 
and $\lambda_2=\pm s_{2}$, respectively. 
\\

The 0-to-2 matrix elements can be derived 
most straightforwardly in the $\Psi^{(*)}$ rest frame ($\Psi$RF) 
(see the kinematics description on the right panel in 
Fig.~\ref{fig:diagram_and_kinematic_configuration_decay}). 
In the following, two combined momenta $p=k_1+k_2$ and $q=k_1-k_2$ 
will be often employed instead of the momenta, $k_1$ and $k_2$, due to their 
symmetric and anti-symmetric properties under the interchange 
$k_1\leftrightarrow k_2$. In the $\Psi$RF, the combined momenta are expressed simply by
\begin{align}
p=\sqrt{p^2}\,(1,\vec 0\,) \quad \mbox{ and }\quad q=\sqrt{p^2}(0,\hat{n}),
\label{eq:combined_momenta}
\end{align}
with the unit vector $\hat{n}=(\sin\theta\cos\phi,\sin\theta\sin\phi,\cos\theta)$ 
given in terms of the polar and azimuthal angles, $\theta$ and $\phi$. It enables 
us to write the momenta of $X_{1}$ and $\bar{X}_2$ as
\begin{align}
k_1=\frac{\sqrt{p^2}}{2}(1,\hat{n}) \quad \mbox{ and }\quad k_2=\frac{\sqrt{p^2}}{2}(1,-\hat{n}),
\label{eq:momenta_k12_in_the_psi_rest_frame}
\end{align}
satisfying $k_1\cdot k_2=g_{\rho\sigma}\,k_1^{\rho}\,k_2^{\sigma}= p^2/2$. 
We adopt the Wick convention~\cite{Wick:1962zz} 
where the helicity states of $X_1$ and $\bar{X}_2$ 
are defined in terms of the rotations, $R(\phi,\theta,0)$ and $R(\pi+\phi,\pi-\theta,0)$, 
respectively. In the $\Psi$RF, then, the polarization vectors of two spin-1 massless 
particles, $X_{1}$ and $\bar{X}_2$, are given by
\begin{align}
\varepsilon(k_1,\pm )
=\frac{1}{\sqrt{2}}(0,\mp \hat{\theta} -i\hat{\phi}) 
\quad \mbox{and} \quad
\varepsilon(k_2,\pm )
=\frac{1}{\sqrt{2}} (0,\mp \hat{\theta} +i\hat{\phi}), 
\label{eq:polarization_vectors}
\end{align}
satisfying $\varepsilon^{\mu}(k_1,\pm)=\varepsilon_{\mu}(k_2,\mp)$ where the two orthonormal unit vectors are 
\begin{align}
\hat{\theta}=(\cos\theta\cos\phi,\cos\theta\sin\phi,-\sin\theta)
\quad\mbox{and}\quad
\hat{\phi}=(\;\,\,-\sin\phi\;\;\,,\;\;\;\,\cos\phi\;\;\;\;,\;\;\;\,\,0\;\;\;\,\,),
\label{eq:theta_and_phi_unit_vectors}
\end{align}
perpendicular to the unit vector $\hat{n}$, respectively. 
The Dirac spinors of two spin-1/2 massless, $X_{1}$ and $\bar{X}_2$, are expressed by
\begin{align}
u( k_{1},\pm\mbox{\small $\frac{1}{2}$})=(p^2)^{\frac14}\,
\left(
\begin{array}{c}
\delta_{-,\pm}\, \xi(\hat{n},-) \\[3mm]
\delta_{+,\pm}\, \xi(\hat{n},+)
\end{array}
\right)
\quad \mbox{and} \quad
v( k_{2},\pm\mbox{\small $\frac{1}{2}$})=-i(p^2)^{\frac14}\,
\left(
\begin{array}{c}
\delta_{+,\pm}\, \xi(\hat{n},+) \\[3mm]
\delta_{-,\pm}\, \xi(\hat{n},-)
\end{array}
\right),
\nonumber
\\
v( k_{1},\pm\mbox{\small $\frac{1}{2}$})=-(p^2)^{\frac14}\,
\left(
\begin{array}{c}
\delta_{+,\pm}\, \xi(\hat{n},-) \\[3mm]
\delta_{-,\pm}\, \xi(\hat{n},+)
\end{array}
\right)
\quad \mbox{and} \quad
u( k_{2},\pm\mbox{\small $\frac{1}{2}$})=i(p^2)^{\frac14}\,
\left(
\begin{array}{c}
\delta_{-,\pm}\, \xi(\hat{n},+) \\[3mm]
\delta_{+,\pm}\, \xi(\hat{n},-)
\end{array}
\right),
\label{eq:dirac_spinors}
\end{align}
satisfying $u(k_1,\pm\frac12)=-i\gamma^0 u(k_2,\mp\frac12)$ and 
$v(k_1,\pm\frac12)=-i\gamma^0 v(k_2,\mp\frac12)$
where the two-component spinors $\xi(\hat{n},\pm)$ are defined as
\begin{align}
\xi(\hat{n},+)= 
\left(
\begin{array}{l}
\cos\frac{\theta}{2}\,e^{-i\frac{\phi}{2}}
\\[2mm]
\sin\frac{\theta}{2}\,e^{i\frac{\phi}{2}}
\end{array}
\right)
\;\; \mbox{ and} \;\;\;\;
\xi(\hat{n},-)= 
\left(
\begin{array}{c}
-\sin\frac{\theta}{2}\,e^{-i\frac{\phi}{2}}
\\[2mm]
\cos\frac{\theta}{2}\,e^{i\frac{\phi}{2}}
\end{array}
\right),
\label{eq:2-component_spinor_wave_functions}
\end{align}
in terms of the polar and azimuthal angles, $\theta$ and $\phi$.

\section{Basic and form-factor operators}
\label{sec:basic_and_form-factor_operators}

The expression of covariant matrix elements in Eq. \eqref{eq:decay_matrix_elements} 
indicates that they can be extracted from the covariant three-point vertices 
directly since all the wave tensors can be derived fully in 
Eqs. \eqref{eq:bosonic_wave_tensor1}, \eqref{eq:bosonic_wave_tensor2} and 
\eqref{eq:fermionic_wave_tensor1}, \eqref{eq:fermionic_wave_tensor2} 
regardless of their interactions. The general covariant three-point vertices can 
be constructed by assembling appropriately their basic building blocks called the 
basic and form-factor operators according to our algorithm. 
Thus, this section focuses on deriving the basic building blocks.
\\

Each of the basic operators can be obtained by 
constructing the covariant three-point vertices for the spin combinations, 
$[J;s_1,s_2]=[0;1,1]$, $[1;1,0]$, $[1;0,1]$, $[2;1,1]$, and $[0;\mbox{$\frac12$},\mbox{$\frac12$}]$, 
$[1;\mbox{$\frac12$},\mbox{$\frac12$}]$. 
A spin-1 massless state is covariant under any Lorentz transformation, but the polarization 
vector, another representation of the state, 
carries an additional term proportional to its momentum under any boost. 
Thus, we introduce the following four dimension-one 
{\it polarization-covariant operators} resulting in the coupled 
polarization vectors to be covariant:
\begin{align}
[ k_1\rho \alpha \mu ]_+
&=i(k_{1\rho} g_{\alpha \mu}-k_{1\alpha}g_{\rho\mu} -k_{1\mu}g_{\rho\alpha}),
\label{eq:polarization-covariant_k1_metric}
\\
[ k_2 \sigma \beta\nu  ]_+
&=i(k_{2\sigma} g_{\beta\nu}  -k_{2\beta} g_{\sigma\nu} - k_{2\nu} g_{\sigma\beta}),
\label{eq:polarization-covariant_k2_metric}
\\
[ k_1\rho \alpha \mu ]_-&=\varepsilon_{\gamma\rho \alpha \mu } \,k^{\gamma}_{1},
\label{eq:polarization-covariant_k1_levi_civita}
\\
[ k_2 \sigma \beta\nu ]_-&=\varepsilon_{\delta \sigma \beta\nu}\, k^{\delta}_{2},
\label{eq:polarization-covariant_k2_levi_civita}
\end{align}
where each operator is the contraction between one 
of the momenta, $k_1$ and $k_2$, and one of the following two fundamental operators,
\begin{align}
[ abcd ]_+ &=i(g_{ab}g_{cd}-g_{ac}g_{bd}-g_{ad}g_{bc}),
\\
[ abcd ]_- &=\varepsilon_{abcd},
\end{align}
defined by metric tensors and a totally antisymmetric Levi-Civita tensor.  
It will be shown that all the basic bosonic operators 
can be constructed by contracting and combining 
the four polarization-covariant operators for which the field descriptions are 
the field-strength tensors and their duals of spin-1 massless $X_1$ and $\bar{X}_2$ 
respectively. On the other hand, the Dirac $u$ and $v$ spinors of a spin-1/2 massless particle 
are covariant themselves with no further adjustments. Thus, the basic fermionic 
operators for the helicity configurations 
$(\lambda_1,\lambda_2)=(\pm \frac12 , \pm \frac12)$ and $(\pm \frac12, \mp \frac12)$, 
will be derived by combining gamma matrices appropriately.
\\

For notational convenience, we introduce a contraction symbol of 
any two fundamental operators:
\begin{align}
[ \wick{a\c1 bcd]_{i}[ e\c1 fgh}]_{j} &=
[ abcd]_i [ efgh]_j\,g^{bf},
\end{align}
in terms of arbitrary indices $(i,j=\pm)$. The full contractions of two 
polarization-covariant operators with identical-sign indices 
are given by the invariant product of two momenta $k_{1,2}$,
\begin{align}
\frac16
[ \wick{k_1\c1 \rho\c2 \alpha\c3 \mu ]_{\pm}[ k_2\c1 \sigma\c2 \beta\c3 \nu} ]_{\pm}
=-k_1\cdot k_2.
\end{align}
Conversely, any full contractions of two fundamental operators with 
opposite-sign indices vanish due to the antisymmetric property of the Levi-Civita tensor. 
In the following, the Levi-Civita tensor is set to follow the 
convention $\varepsilon_{0123}=1$.

\subsection{Basic bosonic scalar operators in the case $[J;s_1,s_2]=[0;1,1]$}

First, we consider the decay of a spin-0 scalar particle $\Psi^{(*)}$ 
into two spin-1 massless particles, $X_{1}$ and $\bar{X}_2$. 
By contracting two polarization-covariant operators with identical-sign indices 
in Eqs. \eqref{eq:polarization-covariant_k1_metric} and 
\eqref{eq:polarization-covariant_k2_metric} or 
in Eqs. \eqref{eq:polarization-covariant_k1_levi_civita} and 
\eqref{eq:polarization-covariant_k2_levi_civita} 
in terms of the four-vector indices, $\rho,\sigma$ and $\mu,\nu$, 
one can find the dimension-two scalar operator yielding 
even-parity scalar matrix elements,
\begin{align}
\frac12 [  \wick{k_1 \c1 \rho  \alpha \c2 \mu  ]_{\pm}
[ k_2 \c1 \sigma  \beta \c2 \nu} ]_{\pm}
\overset{\scriptsize\mbox{eff}}{=}i [ k_1k_2 \alpha\beta ]_{+}
\quad \rightarrow \quad 
\begin{array}{ll}
\Theta^{[0;1,1]}_{ (\pm,\pm)}(k_1,k_2)\!\!\!\!&=
(k_1\cdot k_2)
\\[3pt]
\Theta^{[0;1,1]}_{(\pm,\mp)}(k_1,k_2)\!\!\!\!&=0
\end{array},
\label{eq:even-parity_scalar_contraction}
\end{align}
where the effective equality $\overset{\scriptsize\mbox{eff}}{=}$ 
denotes that the left- and right-hand sides are equal only when 
coupled to the wave tensors of $X_1$ and $\bar{X}_2$, and
$[ k_1k_2 \alpha\beta ]_+=k_1^\rho\, 
k_2^\sigma\, [ \rho\sigma \alpha\beta]_+$ is symmetric 
under the interchanges $\alpha \leftrightarrow \beta$ for four-vector indices 
as well as  $k_{1}\leftrightarrow k_{2}$ for momenta. 
Here, the matrix elements for opposite-sign helicities
vanish due to angular momentum conservation. On the other hand, 
the contraction of two polarization-covariant operators with opposite-sign indices 
in Eqs. \eqref{eq:polarization-covariant_k1_metric} 
and \eqref{eq:polarization-covariant_k2_levi_civita} 
or in Eqs. \eqref{eq:polarization-covariant_k1_levi_civita} 
and \eqref{eq:polarization-covariant_k2_metric} 
yields another scalar operator leading to 
the odd-parity matrix elements,
\begin{align}
\frac12 [  \wick{k_1 \c1 \rho  \alpha \c2 \mu  ]_{\pm}
[ k_2 \c1 \sigma  \beta \c2 \nu} ]_{\mp}
\overset{\scriptsize\mbox{eff}}{=}i[ k_1k_2 \alpha\beta ]_-
\quad \rightarrow \quad
\begin{array}{ll}
\Theta^{[0;1,1]}_{(\pm,\pm)}(k_1,k_2)\!\!\!\!&=
\pm (k_1\cdot k_2)
\\[3pt]
\Theta^{[0;1,1]}_{(\pm,\mp)}(k_1,k_2)\!\!\!\!&=0
\end{array},
\label{eq:odd-parity_scalar_contraction}
\end{align}
in terms of the abbreviation $[ k_1k_2 \alpha\beta ]_-
=k_1^\rho \, k_2^\sigma [ \rho \sigma \alpha\beta ]_-$.
The matrix elements for the opposite-helicity configurations 
in this case vanish due to angular momentum conservation as well. 
Assembling the two scalar operators in Eqs.~\eqref{eq:even-parity_scalar_contraction} 
and \eqref{eq:odd-parity_scalar_contraction} leads to two basic bosonic scalar 
operators $S^{\pm}$:
\begin{align}
&S^{\pm}_{\alpha,\beta}=\frac{i}{2} 
\big([ k_1k_2\alpha\beta]_{+} \pm 
[ k_1k_2\alpha\beta]_-\big)
\quad \rightarrow \quad
\Theta^{[0;1,1]}_{(\lambda_1,\lambda_2)}(k_1,k_2)=
(k_1\cdot k_2)\, \delta_{\lambda_1,\pm} \delta_{\lambda_1,\lambda_2},
\label{eq:basic_bosonic_scalar_operators}
\end{align}
each of which generates a nonzero invariant basic scalar matrix element 
only for one of the identical-sign helicities $(\lambda_1,\lambda_2)=(\pm,\pm)$. 
It indicates that {\it no spin-0 particle can decay into two massless bosons 
with opposite-sign helicities}.

\subsection{Basic bosonic vector operators  in the cases $[J;s_1,s_2]=$ $[1;1,0]$ 
and $[1;0,1]$}

The basic bosonic vector operators are derived in the decays 
of a spin-1 particle $\Psi^{(*)}$ into two massless particles, 
$X_{1}$ and $\bar{X}_2$, of spins 1,0 and 0,1, respectively. 
The proper contractions of the momenta $k_{1,2}$ and 
four polarization-covariant operators 
in Eqs.~\eqref{eq:polarization-covariant_k1_metric} to 
\eqref{eq:polarization-covariant_k2_levi_civita} lead 
to the following even-parity vector matrix elements,
\begin{align}
&i[ k_1k_2 \alpha\mu ]_+
&& \rightarrow &&\,  \Theta^{[1;1,0]}_{(\pm,0)\mu}(k_1,k_2)
= -(k_1\cdot k_2)\,\varepsilon^*_{\bot \mu}(k_1,\pm;k_2),
\label{eq:even_vector_contraction_k1}
\\
\qquad\qquad\qquad\quad
&i[ k_2k_1 \beta\nu ]_+
&& \rightarrow &&\,\Theta^{[1;0,1]}_{(0,\pm)\nu}(k_1,k_2)
= -(k_1\cdot k_2)\,\varepsilon^*_{\bot \nu}(k_2,\pm;k_1),
\qquad\qquad\qquad\quad
\label{eq:even_vector_contraction_k2}
\end{align}
and odd-parity vector matrix elements,
\begin{align}
&i [ k_1k_2 \alpha\mu ]_-
&& \rightarrow &&\,  \Theta^{[1;1,0]}_{(\pm,0)\mu}(k_1,k_2)
= \mp \, (k_1\cdot k_2)\,\varepsilon^*_{\bot \mu}(k_1,\pm;k_2),
\label{eq:odd_vector_contraction_k1}
\\
\qquad\qquad\qquad\quad
&i [ k_2k_1 \beta\nu ]_-
&& \rightarrow &&\,\Theta^{[1;0,1]}_{(0,\pm)\nu}(k_1,k_2)
= \mp \, (k_1\cdot k_2)\,\varepsilon^*_{\bot \nu}(k_2,\pm;k_1),
\qquad\qquad\qquad\quad
\label{eq:odd_vector_contraction_k2}
\end{align}
where the covariant polarization vectors, $\varepsilon_{ \bot}(k_{1,2},\pm;k_{2,1})$, 
orthogonal to the momenta $k_{1,2}$, are expressed by
\begin{align}
&\varepsilon_{\bot }(k_{1,2},\pm;k_{2,1})=\varepsilon(k_{1,2},\pm)
-\bigg(\frac{k_{2,1}\cdot \varepsilon(k_{1,2},\pm)}{k_1\cdot k_2}\bigg) k_{1,2}.
\label{eq:covariant_polarization_vectors_k12}
\end{align}
Combining the vector operators in Eqs. \eqref{eq:even_vector_contraction_k1}, 
\eqref{eq:odd_vector_contraction_k1} and \eqref{eq:even_vector_contraction_k2}, 
\eqref{eq:odd_vector_contraction_k2}, we obtain four dimension-two basic bosonic 
vector operators,
\begin{align}
V^{\pm}_{1\alpha;\mu}&=\frac{i}{2} 
\big([ k_1k_2\alpha\mu ]_+ \pm [  k_1k_2\alpha\mu]_-\big)
&& \rightarrow & 
\Theta^{[1;1,0]}_{(\lambda_1,0)\mu}(k_1,k_2)
&=-(k_1\cdot k_2) \,\varepsilon^*_{\bot \mu}(k_1,\pm;k_2)\, \delta_{\lambda_1,\pm},
\label{eq:basic_bosonic_vector_operators_1}
\\
V^{\pm}_{2\beta;\nu}&=\frac{i}{2} 
\big( [ k_2k_1\beta\nu ]_+ \pm  [  k_2k_1\beta\nu]_-\,\big)
&& \rightarrow & 
\Theta^{[1;0,1]}_{(0,\lambda_2)\nu}(k_1,k_2)
&=-(k_1\cdot k_2) \,\varepsilon^*_{\bot \nu}(k_2,\pm;k_1) \,\delta_{\lambda_2,\pm},
\label{eq:basic_bosonic_vector_operators_2}
\end{align}
each of which generates a nonzero basic vector matrix element only for 
one of the helicity configurations $(\lambda_1,\lambda_2)=(\pm,0)$ or $(0,\pm)$.  
Note that there are only four helicity-specific operators, 
$p_{\mu}S^{\pm}_{\alpha\beta}$ and $q_{\mu}S^{\pm}_{\alpha\beta}$, in 
the covariant three-point vertex for the spin case $[1;1,1]$, 
yielding nonzero matrix elements only for the identical-sign helicities $(\pm,\pm)$ 
due to angular momentum conservation. Thus, the basic 
operators for the opposite-sign helicities must be rank-2 tensors in terms of the $\mu$ index.

\subsection{Basic bosonic tensor operators in the case $[J;s_1,s_2]=$ $[2;1,1]$}

The basic bosonic tensor operators, 
yielding nonvanishing matrix elements only when 
coupled to two polarization vectors of $X_1$ and $\bar{X}_2$ 
with the opposite-sign helicities 
$(\lambda_1,\lambda_2)=(\pm,\mp)$, are constructed in the decay 
of a spin-$2$ particle $\Psi$ into two spin-$1$ massless particles, $X_1$ and $\bar{X}_2$. 
The following even-parity dimension-two contraction of two polarization-covariant operators 
with identical-sign indices for the two four-vector indices, $\rho$ and $\sigma$, 
yields the symmetric tensor matrix elements involving
two covariant polarization vectors 
in Eq.~\eqref{eq:covariant_polarization_vectors_k12},
\begin{align}
&[\wick{k_1 \c1 \rho  \alpha  \mu  ]_+ [ k_2 \c1 \sigma  \beta  \nu} ]_+
-[\wick{k_1 \c1 \rho  \alpha  \mu  ]_- [ k_2 \c1 \sigma  \beta  \nu}]_-
\nonumber
\\
&\rightarrow \quad 
\begin{array}{l}
\Theta^{[2;1,1]}_{(\pm,\mp)\mu\nu}(k_1,k_2)=-(k_1\cdot k_2)
\Big[\varepsilon_{1\bot \mu}^*(k_1,\pm;k_2)\varepsilon_{2\bot \nu}(k_2,\mp;k_1)
+\mu\leftrightarrow \nu\,\Big]
\\[7pt]
\Theta^{[2;1,1]}_{(\pm,\pm)\mu\nu}(k_1,k_2)=0
\end{array},
\label{eq:even-parity_tensor_contraction}
\end{align}
where the notation $\mu\leftrightarrow \nu$ denotes the exchange 
beween two four-vector indices, $\mu$ and $\nu$.
The second term with negative indices 
in Eq.~\eqref{eq:even-parity_tensor_contraction} 
can be rewritten effectively as 
\begin{align}
&[  \wick{k_1 \c1 \rho  \alpha  \mu  ]_-[ k_2 \c1 \sigma  \beta  \nu} ]_-
\overset{\scriptsize\mbox{eff}}{=}
- [\wick{k_1 \c1 \rho  \alpha  \nu ]_+ [ k_2 \c1 \sigma  \beta  \mu}]_+ 
+i [ k_1k_2 \alpha\beta]_+ g_{\mu\nu},
\label{eq:even-parity_identity}
\end{align}
in terms of the first term and even-parity scalar operator. 
The field description for the even-parity tensor operator is 
the energy-momentum tensor of a spin-1 particle.
\\

The odd-parity dimension-two contraction of 
two polarization-covariant operators with opposite-sign indices 
generates the non-vanishing matrix elements only for the opposite-sign 
helicities,  
\begin{align}
&[  \wick{k_1 \c1 \rho  \alpha  \mu  ]_+ [ k_2 \c1 \sigma  \beta  \nu} ]_-
-[ \wick{k_1 \c1 \rho  \alpha  \mu ]_- [ k_2 \c1 \sigma  \beta  \nu} ]_+
\nonumber
\\
&\rightarrow \quad 
\begin{array}{l}
\Theta^{[2;1,1]}_{(\pm,\mp)\mu\nu}(k_1,k_2)=\mp \,(k_1\cdot k_2)
\Big[\varepsilon_{1\bot \mu}^*(k_1,\pm;k_2)\varepsilon_{2\bot \nu}(k_2,\mp;k_1)+\mu\leftrightarrow \nu\,\Big]
\\[7pt]
\Theta^{[2;1,1]}_{(\pm,\pm)\mu\nu}(k_1,k_2)=0
\end{array},
\label{eq:odd-parity_tensor_contraction}
\end{align}
with overall signs depending on the helicities. The fact that no totally antisymmetric rank-5 
tensor exists in four-dimensional spacetime leads to the following identity,
\begin{align}
g_{\nu\mu}\varepsilon_{\alpha\beta\rho\sigma}
-g_{\nu \alpha}\varepsilon_{\mu\beta\rho\sigma}
-g_{\nu \beta}\varepsilon_{\alpha\mu\rho\sigma}
-g_{\nu \rho}\varepsilon_{\alpha\beta\mu\sigma}
-g_{\nu \sigma}\varepsilon_{\alpha\beta\rho\mu}
=0,
\label{eq:schouten_identity}
\end{align}
in terms of metric tensors and Levi-Civita tensors. 
This identity enables us to express the second term 
in the odd-parity tensor operator in Eq.~\eqref{eq:odd-parity_tensor_contraction} 
effectively in terms of the first term and odd-parity scalar operator as 
\begin{align}
[ \wick{k_1 \c1 \rho  \alpha  \mu  ]_- [ k_2 \c1 \sigma  \beta  \nu}]_+
\overset{\scriptsize\mbox{eff}}{=}
-[ \wick{k_1 \c1 \rho  \alpha  \nu  ]_+ [ k_2 \c1 \sigma  \beta  \mu}]_- 
+i [ k_1k_2 \alpha\beta ]_- g_{\mu\nu},
\label{eq:odd-parity_identity}
\end{align}
which takes the same form as the effective identity 
in Eq.~\eqref{eq:even-parity_identity} apart from the signs of indices. 
The basic bosonic tensor operators can be then obtained by
summing and subtracting the two symmetric tensor operators 
in Eqs.~\eqref{eq:even-parity_tensor_contraction} and \eqref{eq:odd-parity_tensor_contraction},
\begin{align}
&T^{\pm}_{\alpha,\beta;\mu\nu}=\frac12\sum_{\tau=\pm}\tau
\Big([\wick{k_1 \c1 \rho  \alpha  \mu  ]_\tau [ k_2 \c1 \sigma  \beta  \nu} ]_\tau
\pm [ \wick{k_1 \c1 \rho  \alpha  \mu  ]_\tau [ k_2 \c1 \sigma  \beta  \nu} ]_{-\tau}\Big)
\nonumber
\\
&\rightarrow \quad 
\Theta^{[2;1,1]}_{(\lambda_1,\lambda_2)\mu\nu}(k_1,k_2)=-(k_1\cdot k_2)
\Big[\varepsilon_{\bot \mu}^*(k_1,\pm;k_2)\varepsilon^*_{\bot \nu}(k_2,\mp;k_1)+\mu\leftrightarrow \nu\,\Big]
\delta_{\lambda_1,\pm}\,\delta_{\lambda_1,-\lambda_2},
\label{eq:basic_bosonic_tensor_operators}
\end{align}
where the basic tensor matrix elements do not vanish only when 
coupled to the two polarization vectors of $X_1$ and $\bar{X}_2$ 
with the opposite helicities $(\pm,\mp)$, respectively.
\\
 
Note that the dimension-four combination of two basic bosonic vector operators 
are equal to the basic bosonic tensor operators with an inner product 
$(k_1\cdot k_2)$ effectively, 
\begin{align}
V^{\pm}_{1\alpha;\mu}V^{\mp}_{2\beta;\nu}
\overset{\scriptsize\mbox{eff}}{=}-
(k_1\cdot k_2)\,T^{\pm}_{\alpha,\beta;\mu\nu}.
\label{eq:fundamental_basic_tensor_operators}
\end{align}
In the previous works~\cite{Choi:2021qsb}, 
the authors have defined all the basic operators to be dimensionless and 
adopted the basic bosonic tensor operators as $
V^{\pm}_{1\alpha;\mu}V^{\mp}_{2\beta;\nu}/
(k_1\cdot k_2)^2$ which can be replaced by the more fundamental structure 
$T^{\pm}_{\alpha,\beta;\mu\nu}/(k_1\cdot k_2)$. 
We notice that the basic bosonic vector operators 
can be obtained by replacing the kinds of indices in the 
scalar operators, i.e. $S^{\pm}_{\alpha,\mu}=V^{\pm}_{1\alpha;\mu}$ 
and $S^{\mp}_{\beta,\nu}=V^{\pm}_{2\beta;\nu}$, Thus, 
all the basic bosonic operators can be obtained through the contractions of the following 
combinations and subtractions of polarization-covariant operators,
\begin{align}
&[ k_1  \rho  \alpha  \mu  ]_+ [ k_2  \sigma  \beta  \nu ]_+
\pm [ k_1  \rho  \alpha  \mu  ]_- [ k_2  \sigma  \beta  \nu ]_-,
\\
&[ k_1  \rho  \alpha  \mu  ]_+ [ k_2  \sigma  \beta  \nu ]_-
\pm [ k_1  \rho  \alpha  \mu ]_- [ k_2  \sigma  \beta  \nu ]_+,
\end{align}
leading to even- and odd-parity matrix elements.

\subsection{Basic fermionic operators in the cases $[J;s_1,s_2]=[0;\frac12,\frac12]$ and 
$[1;\frac12,\frac12]$}

The basic fermionic operators can be obtained in the decays for the cases 
$[J;s_1,s_1]=[0,\frac12,\frac12]$ and $[1;\frac12,\frac12]$. 
Utilizing the explicit expressions of the Dirac $u$ and $v$ spinors 
in Eqs.~\eqref{eq:dirac_spinors}, one can obtain straightforwardly 
the following basic fermionic scalar and vector operators,
\begin{align}
\Sigma^\pm &=\frac{1}{2}(1\mp \gamma_5)
&& \rightarrow &&
\Theta^{[0;\frac12,\frac12]}_{(\lambda_1,\lambda_2)}(k_1,k_2)\,\;=
 i \sqrt{2 k_1\cdot k_2}\;\delta_{\lambda_1,\pm \frac12}\,\delta_{\lambda_1,\lambda_2},
\label{eq:fermionic_scalar_operators} 
\\
\qquad\;\;
\Lambda^\pm_\mu  &= \frac{1}{2}\gamma_{\mu}( 1\pm  \gamma_5)
&& \rightarrow && \Theta^{[1;\frac12,\frac12]}_{(\lambda_1,\lambda_2)\mu}(k_1,k_2)=
-2i\sqrt{k_1\cdot k_2}\;\varepsilon^*_{\bot \mu}(k_1,\pm;k_2)\, 
\delta_{\lambda_1,\pm\frac12}\,\delta_{\lambda_1,-\lambda_2},
\qquad
\label{eq:fermionic_vector_operators}
\end{align}
yielding nonzero basic fermionic matrix elements for the helicity configurations 
$(\lambda_1,\lambda_2)=(\pm\frac12,\pm\frac12)$ and 
$(\pm\frac12,\mp\frac12)$, respectively,
where the basic operators are composed of the Dirac gamma 
matrices $\gamma_{\mu}$ and their product $\gamma_5=i\gamma^0
\gamma^1\gamma^2\gamma^3$. Note that the matrix elements 
in Eqs.~\eqref{eq:fermionic_scalar_operators} 
and \eqref{eq:fermionic_vector_operators} are 
obtained with the minus sign for fermionic cases $(-1)^{4s_1s_2}=-1$ 
appearing due to the configurations of state operators in Eq.~\eqref{eq:decay_matrix_elements}.
\\

{\it Note that the symmetric properties of the covariant tensor currents (for $\mu$ indices) 
and wave tensors of massless particles, $X_1$ and $\bar{X}_2$, 
(for $\alpha$ and $\beta$ indices) allow us to consider only the number of 
basic operators except for the order of four-vector indices 
in the construction of covariant three-point vertices.} 
Thus, the collections of basic bosonic operators 
can be given in the following compact operator form,
\begin{align}
&S^\pm_{\alpha_1,\beta_1}\cdots S^\pm_{\alpha_n\,\beta_n}
&\rightarrow&&&\boldsymbol{S}^{\pm n},
\\[3pt]
&V^\pm_{1\alpha_1;\mu_1}\cdots V^\pm_{1\alpha_n;\mu_n}
&\rightarrow&& &\boldsymbol{V}_1^{\pm n},
\\[3pt]
&V^\pm_{2\beta_1;\mu_1}\cdots V^\pm_{2\beta_n;\mu_n}
&\rightarrow &&&
\boldsymbol{V}_{2}^{\pm n},
\\[3pt]
\qquad\qquad\qquad\qquad
&T^\pm_{\alpha_1,\beta_1;\mu_1\mu_2}\cdots 
T^\pm_{\alpha_n,\beta_n;\mu_{2n-1}\mu_{2n}}
&\rightarrow &&&
\boldsymbol{T}^{\pm n},
\qquad\qquad\qquad\qquad
\end{align}
with the consistent expressions of the fermionic basic operators, 
$\Sigma^{\pm}\rightarrow \boldsymbol{\Sigma}^{\pm}$ and 
$\Lambda^{\pm}_{\mu}\rightarrow \boldsymbol{\Lambda}^{\pm}$, 
in terms of an integer $n$ and $n_{1,2}=s_{1,2}$ or $s_{1,2}-1/2$ 
for integer or half-integer spins $s_{1,2}$ 
where the zeroth and negative $n$th powers of the collected 
basic operators are set to be +1 and 0 respectively. 
These notations will be employed in expressing simply the general 
covariant three-point vertices.

\subsection{Form-factor operators}

Before deriving the general form-factor operators, we discuss one of their simple cases  
for the decay of an off-shell spin-$J$ $\Psi^{*}$ into two on-shell 
spin-0 massless $X_{1}$ and $\bar{X}_2$. In this case, the matrix elements of 
covariant tensor currents are equal to the covariant three-point vertices 
as the wave tensors of $X_{1}$ and $\bar{X}_2$ are given by unity. 
The symmetric property of the covariant matrix elements 
forbids the appearance of Levi-Civita tensors $\varepsilon_{\mu\nu\alpha\beta}$, 
but allows that of two combined momenta, 
$p_{\mu}=(k_1+k_2)_{\mu}$ and $q_{\mu}=(k_1-k_2)_{\mu}$, and 
metric tensors $g_{\mu\nu}$, in the form-factor operators. 
In the $\Psi$RF, $p_{\mu}$ and $g_{\mu\nu}$ are scalars, but 
$q_{\mu}$ is a vector under rotations. It indicates that 
{\it the independent terms in the form-factor operators are the irreducible spherical tensors.} 
It can be checked easily, for example, by observing 
that the four rank-2 tensor operators, 
$q_{\mu_1}q_{\mu_2}$,  $q_{\mu_1}p_{\mu_2}$, 
$p_{\mu_1}p_{\mu_2}$, and $p^2g_{\mu\nu}-p_{\mu}p_{\nu}+q_{\mu}q_{\nu}$ 
are mutually orthogonal due to $q\cdot p=0$ and $q^2=-p^2$ 
in terms of the inner product $A^{\mu_1\mu_2}B_{\mu_1\mu_2}$.
\\

As in the basic operators, the collections of the momenta and metric tensors 
are expressed in a compact operator form~\cite{Choi:2021qsb},
\begin{align}
p_{\mu_1}\cdots p_{\mu_n} \;\rightarrow\; \boldsymbol{p}^n, \;\;\quad
q_{\mu_1}\cdots q_{\mu_n}\;\rightarrow \; \boldsymbol{q}^n,\;\;\quad
g_{\mu_1\mu_2}\cdots g_{\mu_{2n-1}\mu_{2n}} \;\rightarrow\; \boldsymbol{g}^{n},
\label{eq:compact_notation}
\end{align}
in terms of an integer $n$ where the zeroth and negative $n$th 
powers of the operators are set to +1 and 0, respectively. 
Given that the vector operator 
$\boldsymbol{q}$ under rotations plays a role to increase the 
angular momentum, the form-factor operators with the integer-$n$ angular momentum 
can be constructed in an operator form as
\begin{align}
&\big[F^{[J;s_1,s_2]}_{(\lambda s_1,\pm\lambda s_2)}(k_1,k_2)\big]^{n}
_{\mu_1\cdots \mu_{J_{\mp}}}\; \rightarrow \;
\big[\boldsymbol{F}^{[J;s_1,s_2]}_{(\lambda s_1,\pm\lambda s_2)}(k_1,k_2)\big]^{n}
\nonumber
\\
&= \sum_{m=0}^{(J_\mp -n-\eta^{}_{J_\mp-n} )/2} 
A^{[J;s_1,s_2]}_{(\lambda s_1,\pm \lambda s_2)n,m}(k_1,k_2) 
\;\boldsymbol{q}^n 
\boldsymbol{p}^{J_\mp -n-2m}\boldsymbol{g}^{m} \quad \mbox{with} \;\;\lambda=\pm1,
\label{eq:form-factor_operators}
\end{align}
in terms of the form factors $A$ which are the functions of the invariant product 
$k_1\cdot k_2$. Here, the projection operators $
\eta^{}_{J_{\mp}-n}=[1- (-1)^{J_{\mp}-n}]/2$ 
are defined with the abbreviations $J_{\mp}=J-|s_1\mp s_2|$. 
The form-factor operators with a specific helicity configuration 
$(\lambda_1,\lambda_2)$ summed over all the integers $n$ ranging $0\leq n \leq J_{\mp}$ include $\big[(J_{\mp}+2)^2-\eta_{J_{\mp}}^{}]/4$ 
independent terms respectively. 
\\

Let us go back to the simple case with the spin configuration $[J;0,0]$. 
In this case, the covariant matrix elements can be expressed simply by
\begin{align}
\boldsymbol{\Theta}^{[J;0,0]}_{(0,0)}(k_1,k_2)&
=\sum_{n= 0}^{J}\big[\boldsymbol{F}^{[J;0,0]}_{(0,0)}(k_1,k_2)\big]^n,
\label{eq:decay_matrix_elements_00J}
\end{align}
with the 0-to-2 matrix elements given in a compact operator form
$ \Theta_{\mu_1\cdots \mu_J}(k_1,k_2) 
\,\rightarrow \, \boldsymbol{\Theta}_{(k_1,k_2)}$. On the other hand, 
as the matrix elements for an on-shell $\Psi$
are coupled to the $\Psi$ wave tensor, they can be simply given with 
no dependence on the $\boldsymbol{p}$ and $\boldsymbol{g}$ operators 
as
\begin{align}
\boldsymbol{\Theta}^{[J;0,0]}_{(0,0)}(k_1,k_2)&=
\big[\boldsymbol{F}^{[J;0,0]}_{(0,0)}(k_1,k_2)\big]^J=
A^{[J;0,0]}_{(0,0)J,0}\, \boldsymbol{q}^J,
\end{align}
in terms of the form-factor operator with 
a single term including the maximal angular momentum $J$. 
It indicates that the 0-to-2 matrix element for an off-shell $\Psi^*$ 
include additional angular momentum contributions
of which the magnitudes are lower than $J$. For notational convenience, we will omit the 
momentum labels $(k_1,k_2)$ in the form-factor operators and simplify 
the four-vector indices as $\alpha_1\cdots \alpha_{n_1} \rightarrow \alpha$, 
$\beta_1\cdots \beta_{n_2} \rightarrow \beta$, and $\mu_1\cdots \mu_J\rightarrow \mu$, 
unless otherwise specified.
\\

\section{Covariant three-point vertices}
\label{sec:covariant_three-point_vertices}

As mentioned in Sec.~\ref{sec:basic_and_form-factor_operators}, 
the covariant three-point vertices of an integer-spin particle $\Psi$ 
and two massless particles, $X_1$ and $\bar{X}_2$, enable us to obtain 
directly the matrix elements of covariant tensor currents 
because all the explicit expressions of massless wave tensors 
and their properties can be identified fully regardless of their interactions. 
In this section, thus, we show how to construct the three-point vertices 
with the basic and form-factor operators and obtain the selection rules 
for the decay of an off-shell massive particle into two identical massless particles. 

\subsection{Vertices of an off-shell massive and two massless particles}
\label{subsec:vertices_for_an_off_shell_particle}

For a detailed description of the construction, 
we discuss first how to construct a bosonic vertex for a spin configuration $[J;s_1,s_2]$ 
with integer spins $s_{1,2}$ satisfying $s_1\geq s_2>0$.
To generate a non-vanishing matrix element 
for the positive helicities $(+ s_1,+ s_2)$, it is required to 
involve $s_2$ scalar operators $\boldsymbol{S}^{+}$, $s_1-s_2$ 
vector operators $\boldsymbol{V}_{1}^{+}$, and the form-factor operators 
$[\boldsymbol{F}_{(+s_1,+s_2)}]^{n}$ with the nonnegative integers 
$n$ running from 0 to $J-|s_1-s_2|$. In the same manner, 
one can combine the basic operators for the negative helicities 
$(-s_1,-s_2)$ appropriately.
On the other hand, the combination for 
the opposite-sign helicities $(\pm s_1,\mp s_2)$
include the form-factor operators $[\boldsymbol{F}_{(\pm s_1,\mp s_2)}]^{n}$ 
with the nonnegative integers $n$ being from 0 to $J-|s_1+s_2|$. 
\begin{table}[!ht]
\centering
\begin{tabular}{|c|c|c|c|}
\hline
&&&
\\[-5pt]
$n_{\scriptsize\mbox{off}}[J;s_1,s_2]$
& \;\;\;\;\;\;\;$s_1$ and $s_2\neq 0$\;\;\;\;\;\;\; & $s_1$ or $s_2\neq 0$ 
& $s_1$ and $s_2= 0$
\\[7pt]
\hline
&&&
\\[-3pt]
$J < |s_1-s_2|$
& $0$ & 
$0$ & $0$
\\[10pt]
\hline
&&&
\\[-5pt]
\multirow{3}{*}{$|s_1-s_2| \leq J < |s_1+s_2|$}
&$\big[(J_-+2)^2-\eta_{J_-}^{}\big]/2$ & \multirow{3}{*}{$0$} & \multirow{3}{*}{$0$}
\\[8pt]
\cline{2-2}
&&&
\\[-8pt]
&$\boldsymbol{\Theta}_{(\pm,\pm)}$&&
\\[5pt]
\hline
&&&
\\[-6pt]
\multirow{3}{*}{$J \geq |s_1+s_2|$}
& \makecell{$\big[(J_-+2)^2-\eta_{J_-}^{}\big]/2$ \\[3pt] $+\big[(J_++2)^2-\eta_{J_+}^{}\big]/2$} 
& \makecell{$\big[(J_-+2)^2-\eta_{J_-}^{}\big]/2$} & 
\makecell{$\big[(J_-+2)^2-\eta_{J_-}^{}\big]/2$}
\\[14pt]
\cline{2-4}
&&&
\\[-8pt]
&$\boldsymbol{\Theta}_{(\pm,\pm)}$ and $\boldsymbol{\Theta}_{(\pm,\mp)}$
&$\boldsymbol{\Theta}_{(\pm,0)}$ or $\boldsymbol{\Theta}_{(0,\pm)}$
&$\boldsymbol{\Theta}_{(0,0)}$
\\[5pt]
\hline
\end{tabular}
\caption{The numbers of independent terms in the covariant three-point 
vertices for the decays of an off-shell integer spin-$J$ particle $\Psi^*$ 
into two massless particles, $X_1$ and $\bar{X}_2$, of spins $s_{1,2}$ 
which are categorized according 
to three $J$ regions in terms of
the three spin cases, ($s_1$ and $s_2\neq 0$), ($s_1$ or $s_2= 0$), and ($s_1$ and $s_2= 0$). 
$J_\mp$ are the abbreviations of $J-|s_1\mp s_2|$, respectively. 
The even- and odd-$J_{\mp}$ cases are expressed in a unified form 
by introducing of the projection factor $\eta_{n}=[1-(-1)^n]/2$ with an integer $n$.
The corresponding matrix elements are specified under each number.}
\label{table:numbers_off-shell}
\end{table}
Referring to the construction manner of bosonic vertices, one can also 
continue it for the fermionic three-point vertices 
with additional basic fermionic operators. all the numbers of independent terms 
in the covariant three-point vertices for all the spin cases are derived by classifying 
them according to three $J$ regions in terms of
the three spin cases, ($s_1$ and $s_2\neq 0$), ($s_1$ or $s_2= 0$), 
and ($s_1$ and $s_2= 0$), 
in Table.~\ref{table:numbers_off-shell}.\footnote{For notational convenience, 
we take $|s_1+s_2|$ instead of $s_1+s_2$ which are always positive.} 
Note that {\it all the 0-to-2 matrix elements vanish when $J < |s_1-s_2|$ and 
the terms for the opposite-sign helicities can exist only when $J > s_1+s_2$ 
due to angular momentum conservation.}  
\\

We first present the bosonic three-point vertices for 
the decays of an off-shell integer spin-$J$ particle $\Psi^{*}$ into two massless particles, 
$X_1$ and $\bar{X}_2$, of integer spin $s_{1,2}$ , 
\begin{align}
\boldsymbol{\Gamma}^{[J;s_1,s_2]}=&
\sum_{\lambda=\pm}\Bigg\{\theta(J_-)
\sum_{n=0}^{J_-} \big[\boldsymbol{F}^{[J;s_1,s_2]}_{(\lambda s_1,\lambda s_2)}\big]^{n}
\boldsymbol{S}^{\lambda s_{\tiny\mbox{min}}}
+\gamma_{s_{\tiny\mbox{min}}}
\theta(J_+)\sum_{n=0}^{J_+}
\big[\boldsymbol{F}^{[J;s_1,s_2]}_{(\lambda s_1,-\lambda s_2)}\big]^{n}
\boldsymbol{T}^{\lambda s_{\tiny\mbox{min}}}
\Bigg\}
\nonumber
\\
&\times \Big[\boldsymbol{V}_{1}^{\lambda(s_1-s_{\tiny\mbox{min}})}
+\boldsymbol{V}_{2}^{\lambda(s_2-s_{\tiny\mbox{min}})}\Big],
\label{eq:bosonic_vertices}
\end{align}
with the minimum spin symbol 
$s_{\tiny\mbox{min}}=\mbox{min}[s_1,s_2]$, exclusion 
factor $\gamma_{s_{\tiny\mbox{min}}}=1-\delta_{s_{\tiny\mbox{min}},0}$, 
and step functions $\theta(J_{\mp})=1$\, or \,$0$ for $J_{\mp}=J-|s_1\mp s_2| \geq 0$ 
or not respectively. The fermionic three-point vertices 
including the basic fermionic operators, $\mathbf\Sigma^{\pm}$ and 
$\mathbf\Lambda^{\pm}$, as well as basic bosonic operators are given by
\begin{align}
\boldsymbol{\Gamma}^{[J;s_1,s_2]}=&
\sum_{\lambda=\pm}\Bigg\{\theta(J_{-})
\sum_{n=0}^{J_-}
\big[\boldsymbol{F}^{[J;s_1,s_2]}_{(\lambda s_1,\lambda s_2)}\big]^{n}\,
\boldsymbol{\Sigma}^{\lambda}\boldsymbol{S}^{\lambda(s_{\tiny\mbox{min}}-1/2)}
+\theta(J_{+})\sum_{n=0}^{J_+}
\big[\boldsymbol{F}^{[J;s_1,s_2]}_{(\lambda s_1,-\lambda s_2)}\big]^{n}\,
\boldsymbol{\Lambda}^{\lambda}\boldsymbol{T}^{\lambda(s_{\tiny\mbox{min}}-1/2)}
\Bigg\}
\nonumber
\\
&\times \Big[\boldsymbol{V}_{1}^{\lambda(s_1-s_{\tiny\mbox{min}})}
+\boldsymbol{V}_{2}^{\lambda(s_2-s_{\tiny\mbox{min}})}\Big],
\label{eq:fermionic_vertices}
\end{align}
with the minimum spin symbol $s_{\tiny\mbox{min}}=\mbox{min}[s_1,s_2]$ and step 
functions $\theta(J_{\mp})=1$\, or \,$0$ for $J_{\mp}=J-|s_1\mp s_2| \geq 0$ 
or not respectively. The bosonic and fermionic three-point 
vertices expressed in a compact operator form are one of the key results in the present work. 
\\

For analyzing straightforwardly the parity symmetry of the covariant three-point vertices, 
we first discuss how to obtain the vertices for
the even- and odd-parity tensor currents
$\Theta_{\pm\mu}$ of which the space inversions are $ \pm\Theta_{\pm}^{\mu}$ respectively.
In the Wick convention~\cite{Wick:1962zz}, 
the parity transformations of two massless states in the $\Psi$RF are given by 
\begin{align}
|k_1,\lambda_1 \rangle_1 \;\;\rightarrow\;\; \eta^*_{1P}(-1)^{-s_1}
|k_2,-\lambda_1 \rangle_1  
\quad \mbox{and} \quad 
|k_2,\lambda_2 \rangle_{\bar{2}} \;\;\rightarrow\;\; \eta_{2P}^{}(-1)^{-s_2}
|k_1,-\lambda_2 \rangle_{\bar{2}}, 
\end{align}
with the parity phases, $\eta_{1P}^{}$ and $\eta_{2P}^{}$. Then, 
the 0-to-2 matrix elements of the even- and odd-parity currents $\Theta_{\pm\mu}$ 
obey the following parity relations,
\begin{align}
\epsilon_{P} \langle  k_2,-\lambda_1; k_1,-\lambda_2 |\,
\Theta_{\pm}^{\mu}| \,0\, \rangle
&=\pm
\langle  k_1,\lambda_1; k_2,\lambda_2 |\,
\Theta_{\pm \mu}| \,0\, \rangle,
\label{eq:parity_relation}
\end{align}
with the combined phase $\epsilon_P=\eta_{1P}^{}\eta_{2P}^{*}(-1)^{s_1+s_2}$
where the spin labels are omitted in the tensor currents 
for notational convenience. The phase $\epsilon_P$ is set to $(-1)^{4s_1s_2}$ 
because it is valid for all the two identical massless particles regardless 
of their parity phases and spin values as will be discussed 
in Sec~\ref{subsec:vertices_for_two_identical_massless_particles}. Then, 
the above relation can be written in the covariant formulation as
\begin{align}
\varepsilon^{*\alpha}(k_2,-\lambda_1)\;\varepsilon^{*\beta}(k_1,-\lambda_2)\;
\Gamma^{[J;s_1,s_2];\mu}_{\alpha,\beta;\pm}(k_2,k_1)
&=\pm
\varepsilon^{*\alpha}(k_2,\lambda_2)\;\varepsilon^{*\beta}(k_1,\lambda_1)\;
\Gamma^{[J;s_1,s_2]}_{\alpha,\beta;\mu;\pm}(k_1,k_2),
\\
\bar{u}^{\alpha}(k_2,-\lambda_1)\;
\Gamma^{[J;s_1,s_2]}_{\alpha,\beta;\mu;\pm}(k_2,k_1)\;
v^{\beta}(k_1,-\lambda_2)
&=\mp \bar{u}^{\alpha}(k_1,\lambda_1)\;
\Gamma^{[J;s_1,s_2]}_{\alpha,\beta;\mu;\pm}(k_1,k_2)\;
v^{\beta}(k_2,\lambda_2),
\end{align}
for the bosonic and fermionic cases. Employing the relations mentioned in Eqs.~\eqref{eq:polarization_vectors} and~\eqref{eq:dirac_spinors}, 
one can derive the parity relations of the three-point vertices:
\begin{align}
\Gamma^{\alpha,\beta;\mu}_{\pm}(k_2,k_1)=\pm 
\Gamma_{\alpha,\beta;\mu;\pm}(k_1,k_2),
\\
\gamma_0\Gamma^{\alpha,\beta;\mu}_{\pm}(k_2,k_1)\gamma_0=\mp 
\Gamma_{\alpha,\beta;\mu;\pm}(k_1,k_2),
\end{align}
for the bosonic and fermionic cases where we omit the spin labels for simplicity.
Investigating all the basic operators, 
one can find that the superscript signs of all the basic operators are reversed 
through the parity transformation given in the left-hand sides of the above relations. 
Thus, we can conclude that the covariant 
three-point vertices for even- and 
odd-parity tensor currents, respectively, can be constructed by 
imposing the following parity-invariant (PI) conditions only 
on the form-factor operators,
\begin{align}
\big[\boldsymbol{F}_{(\lambda_1,\lambda_2)}^{[J;s_1,s_2]}\big]^n=
\pm (-1)^{4s_1s_2}\big[\boldsymbol{F}_{(-\lambda_1,-\lambda_2)}^{[J;s_1,s_2]}\big]^n
\quad \mbox{for}\;\; \boldsymbol{\Theta}^{[J;s_1,s_2]}_{\pm},
\label{eq:parity_relation_for_form_factor_operators}
\end{align}
in terms of a nonnegative integer $n$. 
Note that the complex form factors in the operators 
$F$ are the functions of the invariant product $(k_1\cdot k_2)$ 
which is invariant under the interchange $k_1 \leftrightarrow k_2$ 
and includes the combined momentum $q_{\mu}$ satisfying $-q^{\mu}=q_{\mu}$ 
in the $\Psi$RF. An interesting observation is that {\it any two spin-0 
massless particles cannot carry both the even- and odd-parity tensor 
currents simultaneously.}

\subsection{Vertices of an on-shell massive and two massless particles}
\label{subsec:vertices_for_an_on_shell_particle}

In the covariant three-point vertices for an on-shell $\Psi$, there are at 
most four independent 
terms as the conditions on the $\Psi$ wave tensor in 
Eqs.~\eqref{eq:symmetric}, \eqref{eq:traceless}, and \eqref{eq:divergence-free} 
forbid the appearance of the momentum $p_{\mu}$ and metric tensor $g_{\mu\nu}$, 
which are scalars under rotations, in the form-factor operators. 
Thus, the form-factor operators are given simply by
\begin{align}
\big[\boldsymbol{F}_{(\lambda s_1, \pm \lambda s_2 )}^{[J;s_1,s_2]}\big]^{J_{\mp}}=
A^{[J;s_1,s_2]}_{(\lambda s_1, \pm \lambda s_2)}\,\boldsymbol{q}^{J_{\mp}} 
\quad \mbox{with} \quad J_{\mp}\geq 0,
\label{eq:on-shell_collection_operators}
\end{align}
in terms of $\lambda=\pm$ and $J_{\mp}=J-|s_1\mp s_2|$. 
The covariant three-point vertices, in this case, can be straightforwardly 
constructed by taking the above form-factor operators into the bosonic 
and fermionic three-point vertices 
for the off-shell cases in Eqs.~\eqref{eq:bosonic_vertices} and \eqref{eq:fermionic_vertices} 
while keeping the other basic operators. 
\begin{table}[!ht]
\centering
\begin{tabular}{|c|c|c|c|}
\hline
&&&
\\[-5pt]
$n_{\scriptsize\mbox{on}}[J;s_1,s_2]$
& \;\;\;\;\;\;\;$s_1$ and $s_2\neq 0$\;\;\;\;\;\;\; & $s_1$ or $s_2\neq 0$ 
& $s_1$ and $s_2= 0$
\\[7pt]
\hline
&&&
\\[-5pt]
$J < |s_1-s_2|$
& $0$ & 
$0$ & $0$
\\[8pt]
\hline
&&&
\\[-7pt]
\multirow{3}{*}{$|s_1-s_2| \leq J < |s_1+s_2|$}
&$2$ & \multirow{3}{*}{$0$} & \multirow{3}{*}{$0$}
\\[4pt]
\cline{2-2}
&&&
\\[-8pt]
&$\boldsymbol{\Theta}_{(\pm,\pm)}$&&
\\[5pt]
\hline
&&&
\\[-8pt]
\multirow{3}{*}{$J \geq s_1+s_2$}
& 4
& 2 
& 1
\\[4pt]
\cline{2-4}
&&&
\\[-8pt]
&$\boldsymbol{\Theta}_{(\pm,\pm)}$ and $\boldsymbol{\Theta}_{(\pm,\mp)}$
&$\boldsymbol{\Theta}_{(\pm,0)}$ or $\boldsymbol{\Theta}_{(0,\pm)}$
&$\boldsymbol{\Theta}_{(0,0)}$
\\[5pt]
\hline
\end{tabular}
\caption{The numbers of independent terms in the covariant three-point vertices 
for the decays of an on-shell integer spin-$J$ particle $\Psi$ 
into two massless particles, $X_1$ and $\bar{X}_2$, of 
integer spins $s_{1,2}$ which are categorized according to 
three $J$ regions in terms of
the three spin cases, ($s_1$ and $s_2\neq 0$), ($s_1$ or $s_2= 0$), and ($s_1$ and $s_2= 0$). 
The corresponding matrix elements are specified under each number.}
\label{table:numbers_on-shell}
\end{table}
It indicates that in an on-shell vertex 
{\it each term for the four helicity configurations involves only 
the maximal angular momentum $J$ 
which is the spin value of the decaying particle $\Psi$.} 
Thus, all the numbers of independent terms in the on-shell three-point vertices can 
be counted simply by classifying them according to three $J$ 
regions in terms of the three cases, ($s_1$ and $s_2\neq 0$), 
($s_1$ or $s_2= 0$), and ($s_1$ and $s_2= 0$), 
as shown in Table.~\ref{table:numbers_on-shell}.

\subsection{Vertices of a massive and two identical massless particles}
\label{subsec:vertices_for_two_identical_massless_particles}

Next, let us consider the decays of an integer spin-$J$ particle $\Psi^{(*)}$ 
into two identical massless particles, $X_1$ and $\bar{X}_2$, of spins $s$ 
satisfying $X_1=\bar{X}_2=X$. In this case, the covariant matrix elements 
obey the following condition,
\begin{align}
{}_{1\bar{2}}\langle k_1,\lambda_1;k_2,\lambda_2|
\Theta^{[J;s,s]}_{\mu_1\cdots \mu_J}|\,0\,\rangle
= {}_{\bar{2}1}\langle k_1,\lambda_1;k_2,\lambda_2|
\Theta^{[J;s,s]}_{\mu_1\cdots \mu_J}|\,0\,\rangle,
\end{align}
in terms of the interachange of the $X_1$ and $\bar{X}_2$ states. 
The Bose/Fermi symmetry allows us to manipulate the condition to be written as
\begin{align}
{}_{1\bar{2}}\langle k_1,\lambda_1;k_2,\lambda_2|
\Theta^{[J;s,s]}_{\mu_1\cdots \mu_J}|\,0\,\rangle
=(-1)^{2s}{}_{1\bar{2}}\langle k_2,\lambda_2;k_1,\lambda_1|
\Theta^{[J;s,s]}_{\mu_1\cdots \mu_J}|\,0\,\rangle,
\end{align}
where the interchange $1\leftrightarrow \bar{2}$ is replaced by 
the interchanges of momenta $k_1\leftrightarrow k_2$ and 
helicities $\lambda_1 \leftrightarrow \lambda_2$. 
The expression in Eq.~\eqref{eq:decay_matrix_elements} then enables 
us to write the above condition in the covariant formulation as
\begin{align}
\varepsilon^{*\alpha}(k_1,\lambda_1)\;\varepsilon^{*\beta}(k_2,\lambda_2)\;
\Gamma^{[J;s,s]}_{\alpha,\beta;\mu}(k_1,k_2)
=
\varepsilon^{*\alpha}(k_2,\lambda_2)\;\varepsilon^{*\beta}(k_1,\lambda_1)\;
\Gamma^{[J;s,s]}_{\alpha,\beta;\mu}(k_2,k_1),
\\
\bar{u}^{\alpha}(k_1,\lambda_1)\;
\Gamma^{[J;s,s]}_{\alpha,\beta;\mu}(k_1,k_2)\;
v^{\beta}(k_2,\lambda_2)
=
-\bar{u}^{\alpha}(k_2,\lambda_2)\;
\Gamma^{[J;s,s]}_{\alpha,\beta;\mu}(k_2,k_1)\;
v^{\beta}(k_1,\lambda_1),
\end{align}
in terms of bosonic and fermionic three-point vertices. By taking 
the replacement $\alpha\leftrightarrow\beta$ with the charge conjugation (only for fermionic cases) $u=C\bar{v}^T$, one can extract the 
following identical-particle (IP) relations~\cite{Choi:2021ewa,Boudjema:1990st} 
of the covariant three-point vertices,
\begin{align}
\qquad\qquad\qquad
\Gamma^{[J;s,s]}_{\alpha,\beta;\mu}(k_1,k_2)
&=\Gamma^{[J;s,s]}_{\beta,\alpha;\mu}(k_2,k_1)&&
\mbox{for bosonic cases},
\qquad\qquad\qquad
\\
\Gamma^{[J;s,s]}_{\alpha,\beta;\mu}(k_1,k_2)
&=C^T\Gamma^{[J;s,s]T}_{\beta,\alpha;\mu}(k_2,k_1)C&&
\mbox{for fermionic cases},
\end{align}
in terms of the charge conjugation operator 
$C=i\gamma^2\gamma^0$ satisfying $C^\dagger=C^T=-C$.
\\

To find the surviving terms in the covariant three-point vertices satisfying the IP relations, 
we first notice that in this case the 
basic bosonic vector operators, $V_1$ and $V_2$, cannot appear 
in the three-point vertices because the two identical massless 
particles of spin $s$ involve only the 
identical-magnitude helicity values. Next, it is necessary to investigate 
how the remaining basic operators are transformed through the replacements, 
$\alpha\rightarrow \beta$ and $k_1\leftrightarrow k_2$, with 
(only for fermionic cases) their charge conjugations. The basic 
bosonic and fermionic scalar operators, $S^{\pm}_{\alpha,\beta}$ 
and $\Sigma^{\pm}$ are invariant under the replacements, but 
the bosonic tensor and fermionic vector operators do not return 
to their original forms:
\begin{align}
T^{\pm}_{\beta,\alpha;\mu\nu}(k_2,k_1)=T^{\mp}_{\alpha,\beta;\mu\nu}(k_1,k_2) 
\quad \mbox{and} \quad
C^T(\Lambda^{\pm}_{\mu})^T C=-\Lambda^{\mp}_{\mu},
\end{align}
where we use the charge conjugations of gamma 
matrices~\cite{Choi:2021ewa} appearing in the fermionic basic operators, 
$\Sigma^{\pm}$ and $\Lambda_{\mu}^{\pm}$,
\begin{align}
C^T \Gamma^T C=\epsilon_C^{}\,\Gamma \;\;\mbox{ with }\;\; \epsilon_C^{}=\left\{
\begin{array}{ll}
+1 & \mbox{for } \Gamma=1,\gamma_5, \mbox{and } \gamma^{\mu}\gamma_5
\\[3pt]
-1 & \mbox{for } \Gamma=\gamma^{\mu}
\end{array}
\right..
\end{align}
In addition, the form-factor operators $\boldsymbol{F}$ with odd numbers 
of the momenta $q_{\mu}$ involve the minus signs with respect to 
the replacement $k_1\leftrightarrow k_2$. The covariant three-point vertices 
then can be given by
\begin{align}
\boldsymbol{\Gamma}^{[J;s,s]}=&
\sum_{\lambda=\pm}\Bigg\{
\sum_{n=0}^J (1-\eta_n) \big[\boldsymbol{F}^{[J;s,s]}_{(\lambda s,\lambda s)}\big]^{n}
\boldsymbol{S}^{\lambda s}
\nonumber
+
\gamma_s\, \theta(J-2s)\sum_{n=0}^{J-2s}
\big[\boldsymbol{F}^{[J;s,s]}_{(\lambda s,-\lambda s)}\big]^n\,
\boldsymbol{T}^{\lambda s}\Bigg\}
\\[3pt]
&\mbox{with}\;\; 
\big[\boldsymbol{F}^{[J;s,s]}_{(+s,-s)}\big]^n=\pm
\big[\boldsymbol{F}^{[J;s,s]}_{(-s,+s)}\big]^n\quad \mbox{for even (+) or odd (-) integers $n$},
\label{eq:off-shell_bosonic_vertices_for_identical_massless_particles}
\end{align}
for bosonic cases with the exclusion factor $\gamma_{s}=1-\delta_{s,0}$ and
\begin{align}
\boldsymbol{\Gamma}^{[J;s,s]}=&
\sum_{\lambda=\pm}\Bigg\{\sum_{n=0}^J 
(1-\eta_n)\big[\boldsymbol{F}^{[J;s,s]}_{(\lambda s,\lambda s)}\big]^{n}\,
\boldsymbol{\Sigma}^{\lambda}
\boldsymbol{S}^{\lambda(s-1/2)}
\nonumber
+
\theta(J-2s)\sum_{n=0}^{J-2s}\big[\boldsymbol{F}^{[J;s,s]}_{(\lambda s,-\lambda s)}\big]^{n}\,
\boldsymbol{\Lambda}^{\lambda}
\boldsymbol{T}^{\lambda(s-1/2)}\Bigg\}
\nonumber
\\[3pt]
&\mbox{with}\;\; 
\big[\boldsymbol{F}^{[J;s,s]}_{(+s,-s)}\big]^n=\mp
\big[\boldsymbol{F}^{[J;s,s]}_{(-s,+s)}\big]^n\quad 
\mbox{for even (-) or odd (+) integers $n$},
\label{eq:off-shell_fermionic_vertices_for_identical_massless_particles}
\end{align}
for fermionic cases in terms of the odd-number projection 
factor $\eta_n=[1-(-1)^n]/2$ with a nonnegative integer $n$. 
\\

As mentioned in Sec.~\ref{subsec:vertices_for_an_on_shell_particle}, 
in the covariant three-point vertices for an on-shell spin-$J$ $\Psi$, 
the term for each helicity configuration involves only the maximal angular momentum $J$. 
Thus, the corresponding three-point vertices are written as
\begin{align}
\boldsymbol{\Gamma}^{[J;s,s]}_{\tiny\mbox{on}}=&
\sum_{\lambda=\pm}\Bigg\{ (1-\eta_J) \big[\boldsymbol{F}^{[J;s,s]}_{(\lambda s,\lambda s)}\big]^{J}
\boldsymbol{S}^{\lambda s}
\nonumber
+
\gamma_s\, \theta(J-2s)\big[\boldsymbol{F}^{[J;s,s]}_{(\lambda s,-\lambda s)}\big]^{J-2s}\,
\boldsymbol{T}^{\lambda s}\Bigg\}
\\[3pt]
&\mbox{with}\;\; 
\big[\boldsymbol{F}^{[J;s,s]}_{(+s,-s)}\big]^{J-2s}=\pm
\big[\boldsymbol{F}^{[J;s,s]}_{(-s,+s)}\big]^{J-2s}\quad 
\mbox{for even (+) or odd (-) integers $J-2s$},
\label{eq:on_shell_bosonic_vertices_for_identical_massless_particles}
\end{align}
and,
\begin{align}
\boldsymbol{\Gamma}^{[J;s,s]}_{\tiny\mbox{on}}=&
\sum_{\lambda=\pm}\Bigg\{
(1-\eta_J)\big[\boldsymbol{F}^{[J;s,s]}_{(\lambda s,\lambda s)}\big]^{J}\,
\boldsymbol{\Sigma}^{\lambda}
\boldsymbol{S}^{\lambda(s-1/2)}
+
\theta(J-2s)\big[\boldsymbol{F}^{[J;s,s]}_{(\lambda s,-\lambda s)}\big]^{J-2s}\,
\boldsymbol{\Lambda}^{\lambda}
\boldsymbol{T}^{\lambda(s-1/2)}\Bigg\}
\nonumber
\\[3pt]
&\mbox{with}\;\; 
\big[\boldsymbol{F}^{[J;s,s]}_{(+s,-s)}\big]^{J-2s}=\mp
\big[\boldsymbol{F}^{[J;s,s]}_{(-s,+s)}\big]^{J-2s}\quad 
\mbox{for even (-) or odd (+) integers $J-2s$},
\label{eq:fermionic_vertices_for_identical_massless_particles}
\end{align}
for bosonic and fermionic cases. Then, 
one can extract immediately a selection rule 
that {\it none of the on-shell odd spin-$J$ massive particle $\Psi$ can decay 
into two identical spin-$s$ massless particles $X$ if $J<2s$~\cite{Choi:2021ewa}.}
\\

In addition, with the PI conditions on the form-factor 
operators in Eq.~\eqref{eq:parity_relation_for_form_factor_operators}, 
the covariant three-point vertices in 
Eqs.~\eqref{eq:off-shell_bosonic_vertices_for_identical_massless_particles} 
and \eqref{eq:off-shell_fermionic_vertices_for_identical_massless_particles} enable us 
to obtain the following selection/exclusion rules of the decays $\Psi^* \rightarrow XX$ 
of an off-shell spin-$J$ massive $\Psi^*$ into two identical spin-$s$ massless particles 
$X$:
\begin{center}
\begin{minipage}{16cm}
\begin{enumerate}
\item No off-shell odd-spin massive particle $\Psi$ can decay into two 
identical spin-$0$ massless particles $X$ if the massless current involves only 
odd angular momenta.
\item Any off-shell spin-1 massive particle can decay only 
into two identical spin-1/2 massless particles carrying the axial vector current when there is no scalar behavior in the massless currents.
\item An off-shell odd-spin massive particle $\Psi$ cannot decay into two 
identical massless bosons carrying only even-parity tensor currents 
if the massless current involves only odd angular momenta.
\item An on-shell odd-spin massive particle $\Psi$ cannot decay into two 
identical massless fermions carrying only odd-parity tensor currents 
if the massless current involves only odd angular momenta.
\end{enumerate}
\end{minipage}
\end{center}
It is an extension to the selection rules~\cite{Choi:2021ewa} 
for the decays of an on-shell particle 
into two identical massless particles of any spin.
Tables.~\ref{table:numbers_off-shell_ip} and \ref{table:numbers_on-shell_ip} 
show the numbers, $n_{\scriptsize\mbox{off-IP}}[J;s,s]$ or 
$n_{\scriptsize\mbox{on-IP}}[J;s,s]$, of independent terms in the 
covariant three-point vertices for the decays of an off-shell or 
on-shell spin-$J$ particle $\Psi^{(*)}$ into two identical spin-$s$ massless 
particles $X$ carrying the even- or odd-parity tensor currents. 
\begin{table}[!ht]
\centering
\begin{tabular}{|c|c|c|}
\hline
&&
\\[-5pt] 
$n_{\scriptsize\mbox{off-IP}}[J;s,s]$ 
& $\boldsymbol{\Theta}_+$ $[\boldsymbol{F}_{(\lambda_1,\lambda_2)}
=(-)^{2s}\boldsymbol{F}_{(-\lambda_1,-\lambda_2)}]$
& $\boldsymbol{\Theta}_-$ $[\boldsymbol{F}_{(\lambda_1,\lambda_2)}
=-(-)^{2s}\boldsymbol{F}_{(-\lambda_1,-\lambda_2)}]$
\\[7pt]
\hline
&&
\\[-9pt]
\multirow{3}{*}{\makecell{$J < 2s\,$ or $\,s=0$\\[5pt]}} 
&\makecell{$(J + 2-\eta_{J}^{})$ \\ $\times (J +4-\eta_{J}^{})/8$} 
&
\makecell{$(J + 2-\eta_{J}^{})$ \\ $\times (J +4-\eta_{J}^{})/8$}
\!\!\!\!\!\!\!\!\!\!\!\!\!\!\!\!\!\!\!\!\!
\makecell{or \, 0} 
\\[12pt]
\cline{2-3}
&&
\\[-11pt]
& $\boldsymbol{\Theta}_{(\pm,\pm)}$ or $\boldsymbol{\Theta}_{(0,0)}$ 
& $\boldsymbol{\Theta}_{(\pm,\pm)}$ or $\boldsymbol{\Theta}_{(0,0)}$
\\[3pt]
\hline
&&
\\[-9pt]
\multirow{3}{*}{\makecell{$J \geq 2s$\\[5pt]}}
&\makecell{$(J+2-\eta_{J}^{})$ \\$ \times (J+4-\eta_{J}^{})/8$
\\[3pt] $+(J-2s+2-\eta_{J-2s}^{})$ \\$ \times (J-2s+4-\eta_{J-2s}^{})/8$}
&\makecell{$(J_-+2-\eta_{J}^{})$ \\$ \times (J_-+4-\eta_{J}^{})/8$
\\[3pt] $+(J-2s+\eta_{J-2s}^{})$ \\$ \times (J-2s+2+\eta_{J-2s}^{})/8$}
\\[25pt]
\cline{2-3}
&\multicolumn{2}{c|}{}
\\[-9pt]
&\multicolumn{2}{c|}{$\boldsymbol{\Theta}_{(\pm,\pm)}$ 
and $\boldsymbol{\Theta}_{(\pm,\mp)}$}
\\[3pt]
\hline
\end{tabular}
\caption{The numbers of independent terms in the covariant three-point 
vertices for the decays of an off-shell integer spin-$J$ particle 
$\Psi^*$ into two identical spin-$s$ massless particles $X$ 
carrying the even- or odd-parity tensor current.
They are classified according to two $J$ regions.   
The odd-number projection factor $\eta_{n}^{}=[1-(-1)^n]/2$ with a 
nonnegative integer $n$ is introduced to describe 
the numbers for the even and odd integer cases of $n=J$ and $J-2s$ collectively. 
Under each number, the corresponding matrix elements are specified.}
\label{table:numbers_off-shell_ip}
\end{table}
\begin{table}[!ht]
\centering
\begin{tabular}{|c|c|c|}
\hline
&&
\\[-5pt] 
$n_{\scriptsize\mbox{on-IP}}[J;s,s]$ 
& $\boldsymbol{\Theta}_+$ 
$[\boldsymbol{F}_{(\lambda_1,\lambda_2)}
=(-)^{2s}\boldsymbol{F}_{(-\lambda_1,-\lambda_2)}]$
& $\boldsymbol{\Theta}_-$ $[\boldsymbol{F}_{(\lambda_1,\lambda_2)}
=-(-)^{2s}\boldsymbol{F}_{(-\lambda_1,-\lambda_2)}]$
\\[7pt]
\hline
&&
\\[-9pt]
\multirow{3}{*}{\makecell{$J < 2s\,$ or $\,s=0$\\[5pt]}} 
&$1-\eta_J^{}$
&$1-\eta_J^{}$\, or\, $0$ 
\\[4pt]
\cline{2-3}
&&
\\[-11pt]
& $\boldsymbol{\Theta}_{(\pm,\pm)}$ or $\boldsymbol{\Theta}_{(0,0)}$ & 
$\boldsymbol{\Theta}_{(\pm,\pm)}$ or $\boldsymbol{\Theta}_{(0,0)}$
\\[3pt]
\hline
&&
\\[-9pt]
\multirow{3}{*}{\makecell{$J \geq 2s$\\[5pt]}}
&$2-\eta_J^{}-\eta_{J-2s}$
&$1-\eta_J^{}+\eta_{J-2s}$
\\[5pt]
\cline{2-3}
&\multicolumn{2}{c|}{}
\\[-9pt]
&\multicolumn{2}{c|}{$\boldsymbol{\Theta}_{(\pm,\pm)}$ 
and $\boldsymbol{\Theta}_{(\pm,\mp)}$}
\\[3pt]
\hline
\end{tabular}
\caption{The numbers of independent terms in the covariant three-point 
vertices for the decays of an on-shell integer spin-$J$ particle 
$\Psi$ into two identical spin-$s$ massless particles $X$, 
which are categorized according to two $J$ regions, 
even- and odd-parity tensor currents. 
The odd-number projection factor $\eta_{n}=[1-(-1)^n]/2$ is introduced 
for describing the numbers collectively. 
Under each number, the corresponding matrix elements are specified.}
\label{table:numbers_on-shell_ip}
\end{table}

\section{Conserved tensor currents of massless particles of any spin}
\label{sec:conserved_tensor_currents}

In this section, all the expressions obtained in constructing 
the 0-to-2 matrix elements of covariant tensor currents
in Sec.~\ref{sec:covariant_three-point_vertices} are employed 
to derive the 1-to-1 matrix elements for scattering processes by utilizing the 
crossing symmetry. Comparing the basic 1-to-1 matrix elements 
for the spacelike and lightlike momentum transfers, 
we show how to identify explicitly the discontinuity of all the 1-to-1 matrix elements.
Allowing the appearance of the noncovariant conserved tensor currents, 
we present how to find explicitly all the conserved tensor currents 
of the massless particles including high spins, enabling us to 
obtain the WW theorem and additional limits on massless particles.

\subsection{Crossing symmetry between the 1-to-1 and 0-to-2 matrix elements}

\begin{figure}[ht!]
\centering
\includegraphics[scale=1.4]{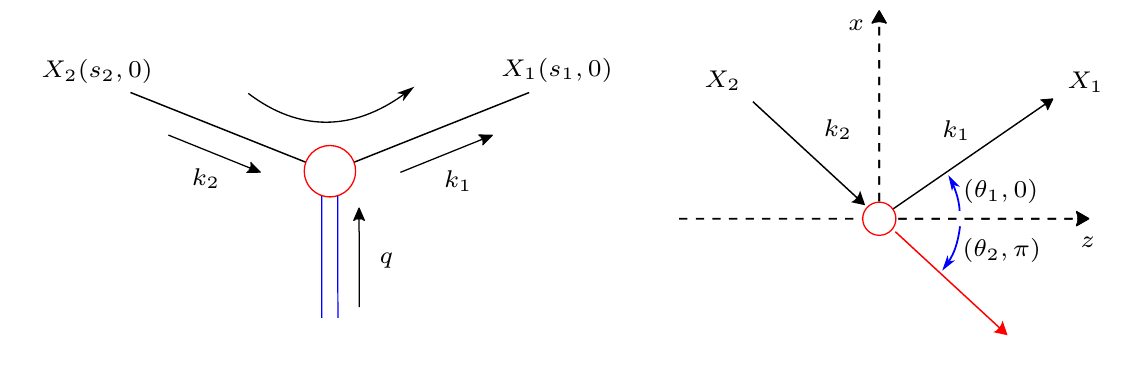}
\caption{A diagram and a kinematic configuration of the scattering processes
for two massless particles, $X_1$ and $X_2$, with the fermion line flowing from left to right. 
(Left) A spin-$s_2$ massless particle $X_2$ with the momentum 
$k_2$ is scattered with an external source, leading to
the production of the spin-$s_1$ massless 
particle $X_1$ with the momentum $k_1$. 
(Right) The scattering processes are dealt with in a reference frame 
with a fixed coordinate system
where $X_1$ and $X_2$ with the different energies
are moving with the polar and azimuthal angles, 
$(\theta_1,0)$ and $(\theta_2,\pi)$, respectively. 
The red arrow is introduced for ease of comparison 
between the momenta $k_{1,2}$.}
\label{fig:diagram_and_kinematic_configuration_scattering}
\end{figure}

The 1-to-1 matrix elements of covariant tensor currents can be obtained 
from the 0-to-2 matrix elements through the crossing symmetry.
According to the configurations of state operators in 
Eq.~\eqref{eq:configuration_of_state_operators},
the 1-to-1 matrix elements do not involve 
the overall sign $(-1)^{4s_1s_2}$ which is even or 
odd for integer or half-integer spin cases. 
Taking the incoming and outgoing particles 
such that the fermion line flows from left to right as described 
on the left panel in Fig.~\ref{fig:diagram_and_kinematic_configuration_scattering},
we can express the 1-to-1 matrix elements as
\begin{align}
&\langle k_1 ,\lambda_1 |\Theta^{[J;s_1,s_2]}_{\mu}| k_2,\lambda_2 \rangle
=\bar{\Theta}^{[J;s_1,s_2]}_{(\lambda_1,\lambda_2)\mu}(k_1,k_2)
\nonumber
\\
&=
\left\{
\begin{array}{ll}
\varepsilon^{*\alpha}(k_1,\lambda_1)\,\bar{\Gamma}^{[J;s_1,s_2]}_{\alpha,\beta;\mu}(k_1,k_2)\,
\varepsilon^{\beta}(k_2,\lambda_2)
&\mbox{for bosonic cases}
\\[3pt]
\bar{u}^{\alpha}(k_1,\lambda_1)\,\bar{\Gamma}^{[J;s_1,s_2]}_{\alpha,\beta;\mu}(k_1,k_2)\,
u^{\beta}(k_2,\lambda_2)
&\mbox{for fermionic cases}
\end{array}
\right.,
\label{eq:scattering_matrix_elements}
\end{align}
with the 1-to-1 covariant three-point vertices 
$\bar{\Gamma}(k_1,k_2)=\Gamma(k_1,-k_2)$ 
defined by the 0-to-2 vertices in Eq.~\eqref{eq:decay_matrix_elements} through the 
replacement $k_2 \rightarrow -k_2$. Connecting the 1-to-1 and 0-to-2 matrix elements 
can be taken by means of the following crossing relations of the $X_2$ 
polarization vectors and Dirac spinors,\footnote{It seems to be 
plausible to define $\varepsilon^*(-k_2,\pm)=-\varepsilon(k_2,\mp)$ 
and $v(-k_2,\pm \mbox{$\frac12$})=u(k_2,\mp\mbox{$\frac12$})$. 
However, it is not essential in obtaining the 
1-to-1 matrix elements, so the study relevant to the extension 
of wave tensors will be reported separately.}
\begin{align}
\varepsilon(k_2,\pm)=-\varepsilon^{*}(k_2,\mp)\quad \mbox{and} \quad 
u(k_2,\pm\mbox{$\frac12$})=-v(k_2,\mp \mbox{$\frac12$}),
\label{eq:crossing_symmetry}
\end{align} 
which can be obtained straightforwardly by referring to their explicit structures in
Eqs.~\eqref{eq:polarization_vectors} and \eqref{eq:dirac_spinors}. 
\\

Note that every bosonic basic operator involves a negative overall sign 
under the replacement $k_2 \rightarrow -k_2$, but the fermionic 
operators are invariant due to the absence of momenta $k_{1,2}$. 
In addition, the replacement leads to the interchange of two combined momenta 
$p_{\mu} \leftrightarrow q_{\mu}$ and so 
the form-factor operators including them in Eq.~\eqref{eq:form-factor_operators}
are replaced by
\begin{align}
&\big[\bar{\boldsymbol{F}}^{[J;s_1,s_2]}_{(\lambda s_1,\mp\lambda s_2)}(k_1,k_2)\big]^{n}
=\big[\boldsymbol{F}^{[J;s_1,s_2]}_{(\lambda s_1,\pm \lambda s_2)}(k_1,-k_2)\big]^{n}
\nonumber
\\
&= \sum_{m=0}^{(J_{\mp}-n-\eta^{}_{J_{\mp}-n} )/2} 
\bar{A}^{[J;s_1,s_2]}_{(\lambda s_1,\mp \lambda s_2);n,m}(k_1,k_2)\;
\boldsymbol{p}^n \boldsymbol{q}^{J_{\mp}-n-2m} \boldsymbol{g}^m,
\end{align}
in terms of the replaced form factors $\bar{A}_{(\lambda s_1, \mp\lambda s_2)}(k_1,k_2)
\equiv A_{(\lambda s_1,\pm \lambda s_2)}(k_1,-k_2)$. Note that the $X_2$ helicity index
gets a minus sign by the crossing relations in Eq.~\eqref{eq:crossing_symmetry}.
For explicit calculations, we present the basic 1-to-1 matrix elements 
for the bosonic and fermionic basic operators:
\begin{align}
\begin{array}{lll}
\bar{S}^{\pm}_{\alpha,\beta}=-S^{\pm}_{\alpha,\beta}& \rightarrow \;\;\;\;&
\bar{\Theta}^{[0;1,1]}_{(\lambda_1,\lambda_2)}(k_1,k_2)=
+(k_1\cdot k_2)\, \delta_{\lambda_1,\pm} \delta_{\lambda_1,-\lambda_2},
\\[7pt]
\bar{V}^{\pm}_{1\alpha;\mu}=-V^{\pm}_{1\alpha;\mu}
& \rightarrow \;\;\;\;& 
\bar{\Theta}^{[1;1,0]}_{(\lambda_1,0)\mu}(k_1,k_2)=
+(k_1\cdot k_2) \,\varepsilon^*_{1\bot \mu}(\pm)\, \delta_{\lambda_1,\pm},
\\[7pt]
\bar{V}^{\pm}_{2\beta;\nu}=-V^{\mp}_{2\beta;\nu}
& \rightarrow \;\;\;\;& 
\bar{\Theta}^{[1;0,1]}_{(0,\lambda_2)\nu}(k_1,k_2)=
-(k_1\cdot k_2) \,\varepsilon_{2\bot \nu}(\pm) \,\delta_{\lambda_2,\pm},
\\[7pt]
\bar{T}^{\pm}_{\alpha,\beta;\mu\nu}=-T^{\pm}_{\alpha,\beta;\mu\nu}& \rightarrow \;\;\;\;&
\bar{\Theta}^{[2;1,1]}_{(\lambda_1,\lambda_2)\mu\nu}(k_1,k_2)=
-(k_1\cdot k_2)\big[\varepsilon^*_{1\bot \mu}(\pm)\varepsilon_{2\bot \nu}^{}(\pm)
+\mu\leftrightarrow \nu\,\big]\, 
\delta_{\lambda_1,\pm} \delta_{\lambda_1,\lambda_2},
\end{array}
\label{eq:scattering_basic_bosonic_operators}
\end{align}
and,
\begin{align}
\begin{array}{lll}
\bar{\Sigma}^\pm =\Sigma^\pm 
& \rightarrow \;\;\;\;&
\bar{\Theta}^{[0;\frac12,\frac12]}_{(\lambda_1,\lambda_2)}(k_1,k_2)\,\;=
i \sqrt{2 k_1\cdot k_2}\;
\delta_{\lambda_1,\pm \frac12}\,\delta_{\lambda_1,-\lambda_2},
\\[5pt]
\bar{\Lambda}^\pm_\mu =\Lambda^\pm_\mu  
& \rightarrow \;\;\;\;&
\bar{\Theta}^{[1;\frac12,\frac12]}_{(\lambda_1,\lambda_2)\mu}(k_1,k_2)=
-2i\sqrt{k_1\cdot k_2}\;\varepsilon^*_{1 \bot \mu}(\pm)\, 
\delta_{\lambda_1,\pm\frac12}\,\delta_{\lambda_1,\lambda_2},
\end{array}
\label{eq:scattering_basic_fermionic_operators}
\end{align}
with the abbreviations $\varepsilon_{1,2\bot}^{}(\pm)=\varepsilon_{\bot}(k_{1,2},\pm;k_{2,1})$. 
Collecting all the aspects of the crossing symmetry, we can observe 
that the diagonal and off-diagonal terms of 0-to-2 matrix elements correspond to 
the off-diagonal and diagonal terms of scattering matrix elements respectively,
\begin{align}
\boldsymbol{\Theta}_{(\lambda_1,\lambda_2)} \leftrightarrow \bar{\boldsymbol{\Theta}}_{(\lambda_1,-\lambda_2)}.
\end{align}
It follows that the numbers of independent terms of the 0-to-2 covariant three-point vertices 
involving the selection rules of the decay process $\Psi \rightarrow XX$ in Tables.~\ref{table:numbers_off-shell}, 
\ref{table:numbers_on-shell}, \ref{table:numbers_off-shell_ip}, 
and \ref{table:numbers_on-shell_ip} can be employed fully in the 
analysis of the 1-to-1 matrix elements. Then, one can recognize straightforwardly that 
{\it two massless particles of spins $s_{1,2}$ cannot carry 
the diagonal scattering matrix elements, i.e. 
$\bar{\boldsymbol{\Theta}}^{[J;s_1,s_2]}_{(\pm s_1,\pm s_2)}=0$, for $J<s_1+s_2$ 
due to angular momentum conservation.}

\subsection{Discontinuity of the 1-to-1 matrix elements}
\label{subsec:discontinuity}

In this section, we discuss how to identify explicitly the discontinuity 
between the 1-to-1 matrix elements 
for the spacelike $(k_1-k_2)^2<0$ and lightlike $(k_1-k_2)^2=0$ cases. 
Since every matrix element can be obtained by combining the basic matrix elements 
and form-factor operators, the task can be focused only on comparing the building blocks in the limiting $(k_1-k_2)^2\rightarrow 0$ and lightlike $(k_1-k_2)^2= 0$ cases. 
\\

For the comparison, we first discuss the limiting case. 
As mentioned before, the covariant matrix elements 
take the same forms in every reference frame, thus, we choose 
a reference frame with a fixed coordinate system where 
their momenta $k_{1,2}$ on the $x$-$z$ plane are given by
\begin{align}
k_1=k^0_1[1,\hat{n}(\theta_1,0)] \quad \mbox{and} \quad k_2=k^0_2[1,\hat{n}(\theta_2,\pi)],
\label{eq:momenta_in_the_limiting_case}
\end{align}
with the unit vector $\hat{n}(\theta,\phi)$ in Eq.~\eqref{eq:combined_momenta}, 
satisfying $k_1\cdot k_2=k^0_1 k^0_2[1-\cos(\theta_1+\theta_2)]$ 
(see the kinematic configuration on the right panel in 
Fig.~\ref{fig:diagram_and_kinematic_configuration_scattering}).
The limits $\theta_{1,2} \rightarrow 0$ on the polar angles then lead 
to the two massless particles to move along the $z$ axis. 
In this frame, the polarization vectors of $X_{1,2}$
are given respectively by
\begin{align}
\varepsilon(k_1,\pm )
=\frac{1}{\sqrt{2}}(0,\mp\cos\theta_1,-i, \pm \sin\theta_1) 
\quad \mbox{and} \quad
\varepsilon(k_2,\pm )
=\frac{1}{\sqrt{2}}(0,\pm\cos\theta_2,+i, \pm \sin\theta_2), 
\label{eq:redefined_polarization_vectors}
\end{align}
satisfying,
\begin{align}
\begin{array}{l}
\displaystyle\lim_{\theta_{1,2}^{} \rightarrow 0}
\sqrt{k_1\cdot k_2}\;\varepsilon^*_{\bot }(k_{1},\pm;k_{2})=\pm k,
\\[10pt]
\displaystyle\lim_{\theta_{1,2}^{} \rightarrow 0}
\sqrt{k_1\cdot k_2}\;\varepsilon_{\bot }(k_{2},\pm;k_{1})=\pm k,
\end{array}
\label{eq:lightlike_limits_of_covariant_polarization_vectors}
\end{align}
in terms of the covariant polarization vectors 
in Eq.~\eqref{eq:covariant_polarization_vectors_k12} 
and the momentum $k=\sqrt{k_1^0k_2^0\,}(1,\hat{z})$ with 
the unit vector $\hat{z}=(0,0,1)$. 
\\

Employing Eqs. \eqref{eq:scattering_basic_bosonic_operators} and 
\eqref{eq:scattering_basic_fermionic_operators} with the covariant 
polarization vectors in Eq.~\eqref{eq:lightlike_limits_of_covariant_polarization_vectors}, 
we obtain the following matrix elements for the basic operators 
in the limiting case,
\begin{align}
\begin{array}{lll}
\bar{S}^{\pm}_{\alpha,\beta}/(k_1\cdot k_2)& \rightarrow \;\;\;\;&
\displaystyle\lim_{\theta_{1,2}^{} \rightarrow 0}
\bar{\Theta}^{[0;1,1]}_{(\lambda_1,\lambda_2)}(k_1,k_2)=
\delta_{\lambda_1,\pm} \delta_{\lambda_1,-\lambda_2},
\\[7pt]
\bar{V}^{\pm}_{1\alpha;\mu}/\sqrt{k_1\cdot k_2}
& \rightarrow \;\;\;\;& 
\displaystyle\lim_{\theta_{1,2}^{} \rightarrow 0}
\bar{\Theta}^{[1;1,0]}_{(\lambda_1,0)\mu}(k_1,k_2)=
\pm k_{\mu}\, \delta_{\lambda_1,\pm},
\\[7pt]
\bar{V}^{\pm}_{2\beta;\nu}/\sqrt{k_1\cdot k_2}
& \rightarrow \;\;\;\;& 
\displaystyle\lim_{\theta_{1,2}^{} \rightarrow 0}
\bar{\Theta}^{[1;0,1]}_{(0,\lambda_2)\nu}(k_1,k_2)=
\mp k_{\nu} \,\delta_{\lambda_2,\pm},
\\[7pt]
\bar{T}^{\pm}_{\alpha,\beta;\mu\nu}& \rightarrow \;\;\;\;&
\displaystyle\lim_{\theta_{1,2}^{} \rightarrow 0}
\bar{\Theta}^{[2;1,1]}_{(\lambda_1,\lambda_2)\mu\nu}(k_1,k_2)=
-2k_{\mu}k_{\nu}\,
\delta_{\lambda_1,\pm} \delta_{\lambda_1,\lambda_2},
\end{array}
\label{eq:basic_bosonic_matrix_elements_in_the_limiting_case}
\end{align}
and,
\begin{align}
\begin{array}{lll}
\bar{\Sigma}^\pm/\sqrt{k_1\cdot k_2}
& \rightarrow \;\;\;\;&
\displaystyle\lim_{\theta_{1,2}^{} \rightarrow 0}
\bar{\Theta}^{[0;\frac12,\frac12]}_{(\lambda_1,\lambda_2)}(k_1,k_2)\,\;=
i \sqrt{2}\;
\delta_{\lambda_1,\pm \frac12}\,\delta_{\lambda_1,-\lambda_2},
\\[5pt]
\bar{\Lambda}^\pm_\mu
& \rightarrow \;\;\;\;&
\displaystyle\lim_{\theta_{1,2}^{} \rightarrow 0}
\bar{\Theta}^{[1;\frac12,\frac12]}_{(\lambda_1,\lambda_2)\mu}(k_1,k_2)=
\mp 2ik_{\mu}\, 
\delta_{\lambda_1,\pm\frac12}\,\delta_{\lambda_1,\lambda_2},
\end{array}
\label{eq:basic_fermionic_matrix_elements_in_the_limiting_case}
\end{align}
with the appropriate normalizations of basic operators such that all of them yield nonzero matrix elements. the divergence-free property of the spin-1 polarization vector 
in Eq.~\eqref{eq:divergence-free} and the equality between the chirality and helicity for massless fermions enforce the basic operators in the lightlike case to be discontinuous 
to the limiting case: 
\begin{align}
\begin{array}{lll}
\bar{S}^{\pm}_{\alpha,\beta}/(k_1\cdot k_2)& \rightarrow \;\;\;\;&
\bar{\Theta}^{[0;1,1]}_{(\lambda_1,\lambda_2)}(\overset{\circ}{k}_1,\overset{\circ}{k}_2)=
\delta_{\lambda_1,\lambda_2},
\\[7pt]
\bar{V}^{\pm}_{1\alpha;\mu}/\sqrt{k_1\cdot k_2}
& \rightarrow \;\;\;\;& 
\bar{\Theta}^{[1;1,0]}_{(\lambda_1,0)\mu}(\overset{\circ}{k}_1,\overset{\circ}{k}_2)=0,
\\[7pt]
\bar{V}^{\pm}_{2\beta;\nu}/\sqrt{k_1\cdot k_2}
& \rightarrow \;\;\;\;& 
\bar{\Theta}^{[1;0,1]}_{(0,\lambda_2)\nu}(\overset{\circ}{k}_1,\overset{\circ}{k}_2)=0,
\\[7pt]
\bar{T}^{\pm}_{\alpha,\beta;\mu\nu}& \rightarrow \;\;\;\;&
\bar{\Theta}^{[2;1,1]}_{(\lambda_1,\lambda_2)\mu\nu}
(\overset{\circ}{k}_1,\overset{\circ}{k}_2)=
2k_{\mu}k_{\nu}\, \delta_{\lambda_1,\lambda_2},
\end{array}
\label{eq:basic_bosonic_matrix_elements_in_the_lightlike_case}
\end{align}
and,
\begin{align}
\begin{array}{lll}
\bar{\Sigma}^\pm/\sqrt{k_1\cdot k_2}
& \rightarrow \;\;\;\;&
\displaystyle\lim_{\theta_{1,2}^{} \rightarrow 0}
\bar{\Theta}^{[0;\frac12,\frac12]}_{(\lambda_1,\lambda_2)}(\overset{\circ}{k}_1,
\overset{\circ}{k}_2)\,\;=0,
\\[5pt]
\bar{\Lambda}^\pm_\mu
& \rightarrow \;\;\;\;&
\displaystyle\lim_{\theta_{1,2}^{} \rightarrow 0}
\bar{\Theta}^{[1;\frac12,\frac12]}_{(\lambda_1,\lambda_2)\mu}
(\overset{\circ}{k}_1,\overset{\circ}{k}_2)=2k_{\mu}\,\delta_{\lambda_1,\lambda_2},
\end{array}
\label{eq:basic_fermionic_matrix_elements_in_the_lightlike_case}
\end{align}
where the momenta $\overset{\circ}{k}_{1,2}=k^0_{1,2}(1,\hat{z})$ are introduced to 
deal with the lightlike cases generically. The differences in overall phases for the 
bosonic tensor and fermionic vector operators are caused by the $z$-axis rotation 
by the angle $\pi$ on the $X_2$ state in Eq.~\eqref{eq:momenta_in_the_limiting_case}, 
implying their continuity. 
\\

Note that the continuity holds for all the terms in the form-factor operators including 
the term $q_{\mu}/\sqrt{k_1\cdot k_2}$ of which the invariance product is given by -1 regardless of the spacelike and lightlike cases. Thus, we can conclude as follows:
\begin{center}
\begin{minipage}{16cm}
\begin{enumerate}
\item all the 1-to-1 matrix elements for the two massless particles with different spins must be discontinuous.
\item The continuity is violated even for the 1-to-1 matrix elements with the opposite-sign helicities of the identical-spin massless particles.
\end{enumerate}
\end{minipage}
\end{center}
An interesting observation is that the basic bosonic scalar operators yield the nonzero 
basic matrix elements for the opposite (spacelike) and identical (lightlike) helicities of 
two massless particles. This is because the momentum-dependent terms are washed out 
due to the divergence-free property of the spin-1 polarization vector as mentioned before 
(see the explicit expressions of bosonic basic operators for the spacelike and lightlike 
cases in Appendix~\ref{sec:explicit_expressions_of_the_bosonic_operators}).

\subsection{1-to-1 matrix elements of conserved tensor currents}

Theoretically, a conserved tensor current $\Theta^{[J;s,s]}_{\mu_1\cdots\mu_J}$ of a spin-$s$ massless particle  is defined to satisfy the following relations,
\begin{align}
\partial^{\mu_i}\Theta_{\mu_1\cdots\mu_i\cdots\mu_J}(x)&=0,
\\
\int d^3\vec x \;\Theta_{0\,\mu_2\cdots\mu_i\cdots\mu_J}(x)
|k,\lambda\rangle &=Q \,k_{\mu_2}\cdots k_{\mu_J}|k,\lambda\rangle, 
\end{align}
in terms of the conserved quantity\,$Q k_{\mu_2}\cdots k_{\mu_J}$ with a real 
constant $Q$. However, one is able to not require the second condition implying the 
continuity of the 1-to-1 matrix elements if it is assumed that a conserved quantity 
is determined by measuring the nearly forward scattering caused by the exchange 
of a spacelike but nearly lightlike massless particle as stated 
in the paper of Weinberg and Witten~\cite{Weinberg:1980kq}. Furthermore, 
the validity of the assumption is checked concretely by employing the physical particle 
state with no definite momentum in Ref.~\cite{Loebbert:2008zz}.
\\

As implied in Sec.~\ref{sec:basic_and_form-factor_operators}, one can construct 
the covariant three-point vertices even yielding the noncovariant matrix elements  
in the covariant formulation. Referring to the discontinuity of matrix elements 
discussed in Sec.~\ref{subsec:discontinuity} 
one can construct all covariant three-point vertices for conserved tensor 
currents. To do so, we define several alternative basic operators as
\begin{align}
\begin{array}{lll}
S_{c\, \alpha,\beta}=-g_{\alpha\beta}& \rightarrow \;\;\;\;&
\bar{\Theta}^{[0;1,1]}_{(\lambda_1,\lambda_2)}(\overset{\circ}{k}_1,\overset{\circ}{k}_2)=
\delta_{\lambda_1,\lambda_2},
\\
T_{c\,\alpha,\beta;\mu\nu}=
[\wick{k_1 \c1 \rho  \alpha  \mu  ]_+ [ k_2 \c1 \sigma  \beta  \nu} ]_+
-[\wick{k_1 \c1 \rho  \alpha  \mu  ]_- [ k_2 \c1 \sigma  \beta  \nu}]_-
& \rightarrow \;\;\;\;&
\bar{\Theta}^{[2;1,1]}_{(\lambda_1,\lambda_2)}(\overset{\circ}{k}_1,\overset{\circ}{k}_2)=
2k_{\mu}k_{\nu}\,\delta_{\lambda_1,\lambda_2},
\end{array}
\end{align}
and, 
\begin{align}
\begin{array}{lll}
\Lambda_{c\,\mu}=\gamma_{\mu}& \rightarrow \;\;\;\;&
\bar{\Theta}^{[1;\frac12,\frac12]}_{(\lambda_1,\lambda_2)}
(\overset{\circ}{k}_1,\overset{\circ}{k}_2)=2 k_{\mu}\, \delta_{\lambda_1,\lambda_2},
\end{array}
\end{align}
yielding nonzero basic matrix elements only for the identical-sign helicities 
in the lightlike case. Note that the scalar operator $S_{c}$ does not involve 
the covariant matrix elements. By collecting the alternative basic and momentum 
$p_{\mu}$ operators appropriately, we can obtain the covariant three-point vertices 
of conserved tensor currents as
\begin{align}
\boldsymbol{\Gamma}^{[J;s,s]}_{\scriptsize\mbox{c}}=
Q\,\boldsymbol{p}^{J-2n}\, \boldsymbol{T}^{n}_c\, 
\boldsymbol{S}^{s-n}_c,
\label{eq:three-point_vertices_for_conserved_tensor_currents_bosonic}
\end{align}
for bosonic cases, and
\begin{align}
\boldsymbol{\Gamma}^{[J;s,s]}_{\scriptsize\mbox{c}}=
Q\,\boldsymbol{p}^{J-2n-1}\,\boldsymbol{\Lambda}_c \,
\boldsymbol{T}^{n}_c\, \boldsymbol{S}^{s-n-1/2}_c,
\label{eq:three-point_vertices_for_conserved_tensor_currents_bosonic}
\end{align}
for fermionic cases with a real constant $Q$ in a compact operator form 
where the nonnegative integer $n$ 
can be chosen arbitrarily under the conditions $J\leq 2n$ and $s\leq n$. 
\\

In order to hold the covariance of conserved tensor currents, 
it is required to set $n=s$ and $n=s-1/2$ for bosonic and fermionic cases respectively 
due to the appearance of the operator $S_c$ involving the noncovariant 
basic matrix elements. It enables us to find the limits
on the conserved current $J_{\mu}$ and energy-momentum tensor $T_{\mu\nu}$ of a massless particle:
\begin{enumerate}
\item The covariance of  $J_{\mu}$ and $T_{\mu\nu}$ of a spin-$s$ massless 
particle holds only for $s\leq 1/2$ and $s \leq 1$ respectively.
\item Oppositely, no noncovariant $J_{\mu}$ and $T_{\mu\nu}$
can exist for $s\leq 1/2$ and $s \leq 1$ respectively.
\item The structures of the noncovariant $J_{\mu}$ 
and $T_{\mu\nu}$ for $s> 1/2$ and $s > 1$ respectively are not uniquely determined.
\end{enumerate}
The first finding is exactly equal to the theorem developed by Weinberg and Witten. 
The second one indicates that 
{\it it is forbidden to construct a theory including spin-$s$ massless particles 
of which charges or energies and momenta 
are not measurable in experiments 
when $s\leq 1/2$ and $s \leq 1$ respectively,} i.e. the charges or energies and momenta of 
such massless particles must have local weights. Contrarily, the third one indicates 
that {\it the spin-$s$ massless particles satisfying $s\leq 1/2$ and $s \leq 1$ 
can be included in a theory only when their charges or energies and momenta 
respectively are not measurable in experiments,} such as the gluon and graviton 
involving the noncovariant conserved current and energy-momentum 
tensor~\cite{Loebbert:2008zz}.

\section{Conclusion}
\label{sec:conclusion}

We have presented an efficient algorithm to construct all the matrix elements 
of covariant tensor currents of massless particles of any spins 
in the covariant formulation. The matrix elements are expressed in a helicity basis 
adopting the Wick convention. The construction has been 
taken first in the two-body decay of an integer spin-$J$ particle $\Psi^{(*)}$ 
into two massless particles of arbitrary spins $s_{1,2}$. 
To cover the off-shell $\Psi^*$ decay, we have taken only the least 
constraint on the covariant tensor currents which is the symmetricity 
of the tensor currents.
\\

After deriving all the basic and form-factor operators which are the 
basic building blocks involving basic matrix elements, we have constructed 
the general three-point vertices of an off-shell particle and two massless 
particles straightforwardly by assembling the building blocks according to 
our algorithm. All the matrix elements can be extracted directly from the 
covariant three-point vertices as the wave tensors 
of massless particles are determined regardless of their interactions.
In the construction of basic operators, we have 
found more fundamental structures of basic bosonic tensor operators than those given in the prototype~\cite{Choi:2021qsb} of our algorithm. The revised basic operators would be 
more useful in extending the algorithm to cover more general event topologies. 
\\

Employing the constructed three-point vertices, we have shown 
the selection rules for the decay of an off-shell massive particle into two identical 
massless particles of any spin, which are {\it the generalization} of the Landau-Yang theorem.
The 1-to-1 matrix elements were derived from the 0-to-2 matrix elements 
by utilizing the crossing symmetry. Taking the lightlike limit on the matrix elements 
with the spacelike momentum transfer, we have investigated the discontinuity of 
all the basic matrix elements, resulting in the full identification of the discontinuity 
of all the matrix elements. Finally, we have constructed 
the covariant three-point vertices for all the conserved tensor currents with the 
introduction of several alternative basic operators. From the vertices, 
we have extracted the WW theorem directly with {\it additional limits} 
on massless particles including high spins. 
\\

As a natural extension of this work, the development of an algorithm for constructing 
all the covariant three-point vertices for all the off-shell particles is under study at present. 
In addition, the construction of all the four-point vertices will be reported separately. 
After finishing the extensions, a program to generate all the three- and 
four-point vertices and the corresponding Lagrangian operators 
will be presented for ease of usage of the algorithm. All of these extensions will enable 
us to deal with straightforwardly the phenomenological and theoretical aspects of 
exotic interactions of particles including high spins.

\section*{Acknowledgments}

I thank Prof. S. Y. Choi for the active discussion on this work.
This work has been supported by a KIAS Individual Grant (QP090001) via 
the Quantum Universe Center at Korea Institute for Advanced Study.

\appendix

\section{Explicit expressions of the bosonic basic operators}
\label{sec:explicit_expressions}

In this section, we present the explicit expressions of all the bosonic basic operators for the 
spacelike and lightlike momentum transfers.

\subsection{Spacelike case $(k_1-k_2)^2<0$}
\label{sec:explicit_forms_in_the_spacelike_case}

Employing the polarization-covariant operators in 
Eqs.~\eqref{eq:polarization-covariant_k1_metric} 
to~\eqref{eq:polarization-covariant_k2_levi_civita}, 
we can calculate the basic scalar and vector operators as
\begin{align}
S^{\pm}_{\alpha\beta}&\;\overset{\scriptsize\mbox{eff}}{=}\;\frac12
\Big[
-(k_1\cdot k_2)g_{\alpha\beta}+k_{2\alpha}k_{1\beta}
\pm ik_1^{\rho}k_2^{\sigma}\varepsilon_{\rho\sigma \alpha\beta}\Big],
\\[5pt]
V^{\pm}_{1\alpha\mu}&\;\overset{\scriptsize\mbox{eff}}{=}\;\frac12
\Big[
-(k_1\cdot k_2)g_{\alpha\mu}+k_{2\alpha}k_{1\mu}
\pm ik_1^{\rho}k_2^{\sigma}\varepsilon_{\rho\sigma \alpha\mu}\Big],
\\[5pt]
V^{\pm}_{2\beta\mu}&\;\overset{\scriptsize\mbox{eff}}{=}\;\frac12
\Big[
-(k_1\cdot k_2)g_{\beta\nu}+k_{1\beta}k_{2\mu}
\pm ik_2^{\sigma}k_1^{\rho}\varepsilon_{\sigma\rho\beta\mu}\Big],
\end{align}
with the absence of the terms involving $k_{1\alpha}$ or $k_{2\beta}$ due 
to the divergence-free property of polarization vectors. 
Employing the effective identities in Eqs.~\eqref{eq:even-parity_identity} 
and~\eqref{eq:odd-parity_identity}, the bosonic tensor
operators can be written effectively as
\begin{align}
T^{\pm}_{\alpha\beta;\mu\nu}&\overset{\scriptsize\mbox{eff}}{=}\;
\frac12 \bigg\{
\Big[-(k_1\cdot k_2)g_{\alpha\mu}g_{\beta\nu}
+g_{\alpha\mu}k_{1\beta}k_{2\nu}+g_{\beta\nu}k_{2\alpha}k_{1\mu}
-k_{1\mu}k_{2\nu}g_{\alpha\beta}-\mu\leftrightarrow \nu\Big]
\nonumber
\\
&-g_{\mu\nu}\Big[(k_1\cdot k_2)g_{\alpha\beta}-k_{2\alpha}k_{1\beta}\Big]
\pm 
i\Big[g_{\alpha\mu}k^{\sigma}_{2}k^{\rho}_{1}\varepsilon_{\sigma \rho \beta\nu }
-k_{1\mu} k^\sigma_2 \varepsilon_{\sigma \alpha\beta\nu}+\mu\leftrightarrow \nu\Big]
\pm 
ig_{\mu\nu}k_1^{\rho}k_2^{\sigma}\varepsilon_{\rho\sigma \alpha\beta}\bigg\},
\end{align}
with the exchange $\mu \leftrightarrow \nu$ between two four-vector indices, $\mu$ and $\nu$.

\subsection{Lightlike case $(k_1-k_2)^2=0$}
\label{sec:explicit_forms_in_the_lightlike_case}

In the lightlike case with the momenta $\overset{\circ}{k}_{1,2}=k^0_{1,2}(1,\hat{z})$, 
all the odd-parity contractions vanish due to the antisymmetric property 
of the Levi-Civita tensor. In addition, the divergence-free property of polarization 
vectors removes all the momenta involving the four-factor indices, $\alpha$ and $\beta$, 
in the basic operators. Thus, the basic operators can be 
effectively given by

\begin{align}
S^{\pm}_{\alpha\beta}/
(\overset{\circ}{k}_1\cdot \overset{\circ}{k}_2)
\overset{\scriptsize\mbox{eff}}{=}-g_{\alpha\beta},
\quad \;\;
V^{\pm}_{1\alpha\mu}/
(\overset{\circ}{k}_1\cdot \overset{\circ}{k}_2)
\overset{\scriptsize\mbox{eff}}{=}-g_{\alpha\mu},
\quad \;\;
V^{\pm}_{2\beta\mu}/
(\overset{\circ}{k}_1\cdot \overset{\circ}{k}_2)
\overset{\scriptsize\mbox{eff}}{=}-g_{\beta\mu},
\end{align}
and,
\begin{align}
T^{\pm}_{\alpha\beta;\mu\nu}
\overset{\scriptsize\mbox{eff}}{=}
-k_{\mu}k_{\nu} g_{\alpha\beta}.
\end{align}
with the momentum $k=\sqrt{k_1^0k_2^0}(1,\hat{z})$ and the invariant product 
$(\overset{\circ}{k}_1\cdot \overset{\circ}{k}_2)$ introduced 
for comparing the limiting case $(k_1-k_2)^2\rightarrow 0$ 
and the lightlike case $(k_1-k_2)^2=0$ 
in Sec.~\ref{subsec:discontinuity}.
.

\end{document}